\documentclass[traditabstract]{aa}  
\usepackage{graphicx}
\usepackage{txfonts}
\begin{document}

   \title{Two-dimensional simulations of mixing in classical novae: the effect of the white dwarf composition and mass}

   \author{Jordi Casanova
          \inst{1}
          \and
          Jordi Jos\'{e}
          \inst{2,3}
          \and
          Steven N. Shore\inst{4}
          }

   \institute{Physics Division, Oak Ridge National Laboratory, PO Box 2008, Oak Ridge, TN 37831-6354, USA
              \email{novaj@ornl.gov}
              \and
              Departament de F\'{i}sica, EEBE, Universitat Polit\`{e}cnica de Catalunya, c/Eduard Maristany 10, E-08930 Barcelona, Spain
              \and
              Institut d'Estudis Espacials de Catalunya, c/Gran Capit\`{a} 2-4, Ed. Nexus-201, E-08034 Barcelona, Spain
         \and
             Dipartimento di F\'{i}sica "Enrico Fermi",  Universit\`{a} di Pisa and INFN, Sezione di Pisa, Largo B. Pontecorvo 3, 56127 Pisa, Italy\\
             }


 
  \abstract
  {Classical novae are explosive phenomena that take place in stellar binary systems. They are powered by mass transfer from a low-mass main sequence star onto a white dwarf (either CO or ONe). The material accumulates for $10^{4}-10^{5}$ yr until ignition under degenerate conditions, resulting in a thermonuclear runaway. 
The nuclear energy released produces peak temperatures 
of $\sim 0.1-0.4$ GK. During these events, $10^{-7}-10^{-3} M_{\odot}$ enriched in intermediate-mass elements (with respect to solar abundances) are ejected into the interstellar medium. However, the origin of the large metallicity enhancements and the inhomogeneous distribution of chemical species observed in high-resolution spectra of ejected nova shells is not fully understood.}
   {Recent multidimensional simulations have demonstrated that Kelvin-Helmholtz instabilities that operate at the core-envelope interface can naturally produce self-enrichment of the accreted envelope with material from the underlying white dwarf at levels that agree with observations. However, such multidimensional simulations have been performed for a small number of cases, and much of the parameter space remains unexplored.}
   {Here we investigate the dredge-up, driven by Kelvin-Helmholtz instabilities, for white dwarf masses in the range 0.8-1.25 $M_{\odot}$ and different core compositions (CO-rich and ONe-rich substrates). We present a set of five numerical simulations performed in two dimensions aimed at analyzing the possible impact of the white dwarf mass (and composition) on the metallicity enhancement and on the explosion characteristics.}
   {We observe greater mixing ($\sim$ 30$\%$ higher when measured in the same conditions), at the time we stop the simulations, and more energetic outbursts for ONe-rich substrates than for CO-rich substrates and for more massive white dwarfs.}
   {}

   \titlerunning{2D simulations of mixing in novae}

   \maketitle
%

\section{Introduction}

 Classical novae are stellar explosions powered by thermonuclear runaways (TNR) on the surface of a white dwarf after an accretion episode of hydrogen-rich matter from the companion star at typical rates $\sim$ $10^{-10}$ to $10^{-9}$ $M_{\odot}$ yr$^{-1}$.  The TNR drives peak temperatures of $\sim$ (1 - 4) $\times$ $10^{8}$ K. Between $10^{-7}$ and $10^{-3}$ $M_{\odot}$ are ejected with peak velocities that achieve several thousand km s$^{-1}$. The suite of nuclear processes that operate in the envelope results in non-solar isotopic abundance ratios in the ejecta \citep{Gehrz98,Dow13,Kelly13}. Classical novae are spectroscopically classified as neon novae (characterized by the presence of intense Ne lines; these novae are thought to take place on ONe white dwarfs) and non-neon novae (where the Ne lines are absent; they take place on CO white dwarfs). They contribute to the production of Galactic $^{15}$N, $^{17}$O, and $^{13}$C \citep{{Star08,Star16},Jose98,Jose16}. Other species, such as $^{31}$P, $^{32}$S, and $^{35}$Cl can also be produced in the most massive ONe novae, since the pressure achieved at the base of the accreted envelope is higher in these novae. In ONe novae, this translates into higher peak temperatures and nuclear processing that extends towards heavier isotopes, beyond the CNO region \citep{Jose16,Jose08b}.
  
While the matter accreted from the companion onto the white dwarf has approximately solar composition (Z $\sim$ 0.02), the ejecta does not. The metallicity enhancements inferred from observations reveal values in a wide range from slightly above solar to Z $\sim$ 0.80\footnote{E.g., V1370 Aql 1982 \citep{Sni87,And94}, although it is worth mentioning that these studies used ionization correction methods.} for the more massive ONe novae (see \citet{Van97} and \citet{Gehrz98}, and references therein). Due to the moderately low peak temperatures achieved during the outburst, nuclear processing in the accreted envelope alone is unlikely to account for the high metallicities inferred from observations. Instead, mixing at the core-envelope interface is a more reliable alternative to explain the metallicity enhancement in the ejecta. One-dimensional (spherical symmetric) models often assume mixing, at a constant rate, between the material transferred from the companion and the outermost layers of the underlying white dwarf, whose composition reflects its previous evolutionary history. Such mixing is, however, most likely time dependant. For the 1D simulations, several mixing mechanisms have been proposed, such as diffusive mixing \citep{Prialnik84,Kov85,Fuji92,{Iben91,Iben92}} and shear mixing \citep{Durisen77,Kippen78,MacDon83,Livio87,Fuji88,Sparks87,{Kutter87,Kutter89}}. Yet none can reproduce the full range of metallicity enhancements inferred from observations \citep{Livio90}. Aside from the difficulties faced by 1D models in the search for a feasible mixing mechanism, spherically symmetric simulations cannot explore how the ignition begins and how the deflagration spreads throughout the accreted envelope \citep{Shara82}. Therefore, multidimensional hydrodynamical calculations are required to shed more light on the unexplained features of classical nova explosions, in particular, the role of fluid instabilities for the CNO enhancement. 

Recent studies in 2-D and 3-D simulations (see \citet{Glasner12,Cas16}, and references therein) confirmed that Kelvin-Helmholtz instabilities at the core-envelope interface can dredge up material from the underlying white dwarf and efficiently enrich the accreted envelope to levels in agreement with observations. While most of the  multidimensional nova studies performed to date have focused on mixing with CO-rich substrates, two independent efforts have analyzed the effect of different compositions for the underlying white dwarf \citep{{Glasner12,Glasner14},Cas16}. The 2-D models by \citet{Glasner12,Glasner14} assumed mass-accretion of solar composition onto a 1.147 $M_{\odot}$ white dwarf with a maximum resolution of 1.4 $\times$ 1.4 km$^{2}$, in spherical-polar coordinates. The authors found that Kelvin-Helmholtz instabilities can enrich the accreted envelope with material from the underlying white dwarf, independent of the nature of the chemical substrate (CO, pure He, pure $^{16}$O, and pure $^{24}$Mg), but the study did not consider a realistic composition of an ONe white dwarf. The 3-D work in cartesian coordinates by \citet{Cas16}, employed instead an ONe-rich substrate model based on \citet{Ritossa96}, assuming accretion of solar composition onto a 1.25 $M_{\odot}$ ONe white dwarf with a maximum resolution of 1.56 $\times$ 1.56 $\times$ 1.56 km$^{3}$. The authors found larger metallicity enhancements and longer durations of the thermonuclear runaway for ONe-rich substrates, but further analysis of the influence of the core composition for different white dwarf masses is needed. In this paper, we present a 2-D study of mixing at the core-envelope interface (CEI) during nova outbursts with self-consistent models to explore the effect of the white dwarf mass and composition, aimed at understanding the impact of the white dwarf gravity for a given chemical composition.

\section{Input physics and initial setup}

   We used the 1D implicit Lagrangian hydrodynamic code SHIVA to simulate the accretion of solar composition material (Z=0.02) onto the white dwarf at a rate of 2 $\times$ 10$^{-10}$ $M_{\odot}$ yr$^{-1}$ \citep{Jose98,Jose16}. We assumed cold accretion with no pre-mixing, at a constant rate. For this study, we have computed three models of CO-rich and two models of ONe-rich substrates with different white dwarf masses: 0.8 $M_{\odot}$ CO, 1.0 $M_{\odot}$ CO, 1.15 $M_{\odot}$ CO, 1.15 $M_{\odot}$ ONe, and 1.25 $M_{\odot}$ ONe. 
For CO-rich substrate models, the composition of the underlying white dwarf is X($^{12}$C) = X($^{16}$O) = 0.5. For ONe-rich substrate models, we adopted the composition of the outer white dwarf layers from \citet{Ritossa96}: X($^{16}$O) = 0.511, X($^{20}$Ne) = 0.313, X($^{23}$Na) = 0.0644, X($^{24}$Mg) = 0.0548, X($^{25}$Mg) = 0.0158, X($^{27}$Al) = 0.0108, X($^{12}$C) = 0.00916, X($^{26}$Mg) = 0.00989, X($^{21}$Ne) = 0.00598, and X($^{22}$Ne) = 0.00431.

The structures were mapped onto a two-dimensional cartesian grid when the CEI temperature reached T=0.1 GK, and the subsequent evolution was followed with the multidimensional parallelized explicit Eulerian FLASH code. The FLASH code is based on the piecewise parabolic interpolation of physical quantities to solve the hydrodynamic equations that describe the stellar plasma, with a timestep limited by the Courant-Friedrichs-Lewy condition, and uses adaptive mesh refinement to resolve critical features along the computational domain \citep{Fryxell00}. Thermal diffusion is implemented by adding the heat flux in the energy equation which is calculated using a Rossland mean opacity that includes radiative and conductive contributions \citep{Timmes00}. The physical description of the fluid is closed by the Helmholtz equation of state, that is suitable for degenerate stellar matter and includes contributions from ions treated as an ideal gas, radiation as a blackbody, and degenerate electrons and positrons described with a non-interacting Fermi gas \citep{Timmes99,Timmes00b}. FLASH uses dissipation algorithms, as described in \cite{Col84}, such as a flattening procedure and a monotonicity constraint (rather than artificial viscosity) to control oscillations near discontinuities, a feature shared with the MUSCL scheme of van Leer \citep{Leer79}. The calculations were performed on the MareNostrum supercomputer (Barcelona Supercomputing Center).

The two-dimensional computational domain for the CO nova models is $800\times800$ km$^{2}$, initially comprising 128 unevenly spaced radial layers, and 1024 equally spaced grid points along the horizontal axis. For the ONe nova models, the two-dimensional computational domain is $800\times400$ km$^{2}$, with 96 unevenly spaced radial zones and 1024 equally spaced zones along the horizontal axis. The initial spacing is determined by the refinement criterion used by the FLASH code, which tends to refine the regions characterized by rapidly changing physical variables (i.e., the CEI rather than other regions of the accreted envelope).
The critical mass and extent of the accreted envelopes depend sensitively on the properties of the underlying white dwarf. The envelope is more massive and larger for CO-rich than for the more massive ONe-rich white dwarfs, although the latter reach higher densities because of the white dwarf mass-radius relation \citep{Star98,Jose98,Yaron05,Jose16}.
A maximum resolution of $0.78\times0.78$ km$^{2}$ was adopted to capture the mixing process operating at the CEI in all the simulations reported in this paper. Initially, the structure was forced to be in hydrostatic equilibrium throughout the envelope. This condition is reinforced with a reflecting boundary condition at the bottom (i.e., the sign of the velocity is reversed), and an outflow condition at the top (which allows matter to escape). Periodic lateral boundary conditions were adopted \citep{Zingale02,{Cas10,Cas11a,Cas11b,Cas16}}.

We employed a reduced nuclear network containing 13 chemical species ($^{1}$H, $^{4}$He, $^{12,13}$C, $^{13,14,15}$N, $^{14,15,16,17}$O, and $^{17,18}$F) linked through 18 nuclear reactions, mainly p-captures and $\beta^{+}$-disintegrations, to compute the energetics of CO-rich novae \citep{Cas10,Cas11a,Cas11b}. For ONe-rich novae, we employed an extended nuclear network containing 31 chemical species ($^{1}$H, $^{4}$He, $^{12,13}$C, $^{13,14,15}$N, $^{14,15,16,17}$O, $^{17,18}$F, $^{20,21}$Ne, $^{21,22,23}$Na, $^{22,23,24,25,26}$Mg, $^{24,25,26g,26m,27}$Al, and $^{26,27,28}$Si), since the main nuclear processing extends beyond the CNO cycle in this type of novae \citep{Cas16}. The reaction rates were taken from STARLIB nuclear reaction library \citep{Iliadis10}.   

   \begin{figure*}
   \includegraphics[width=0.475\textwidth,height=0.3\textheight]{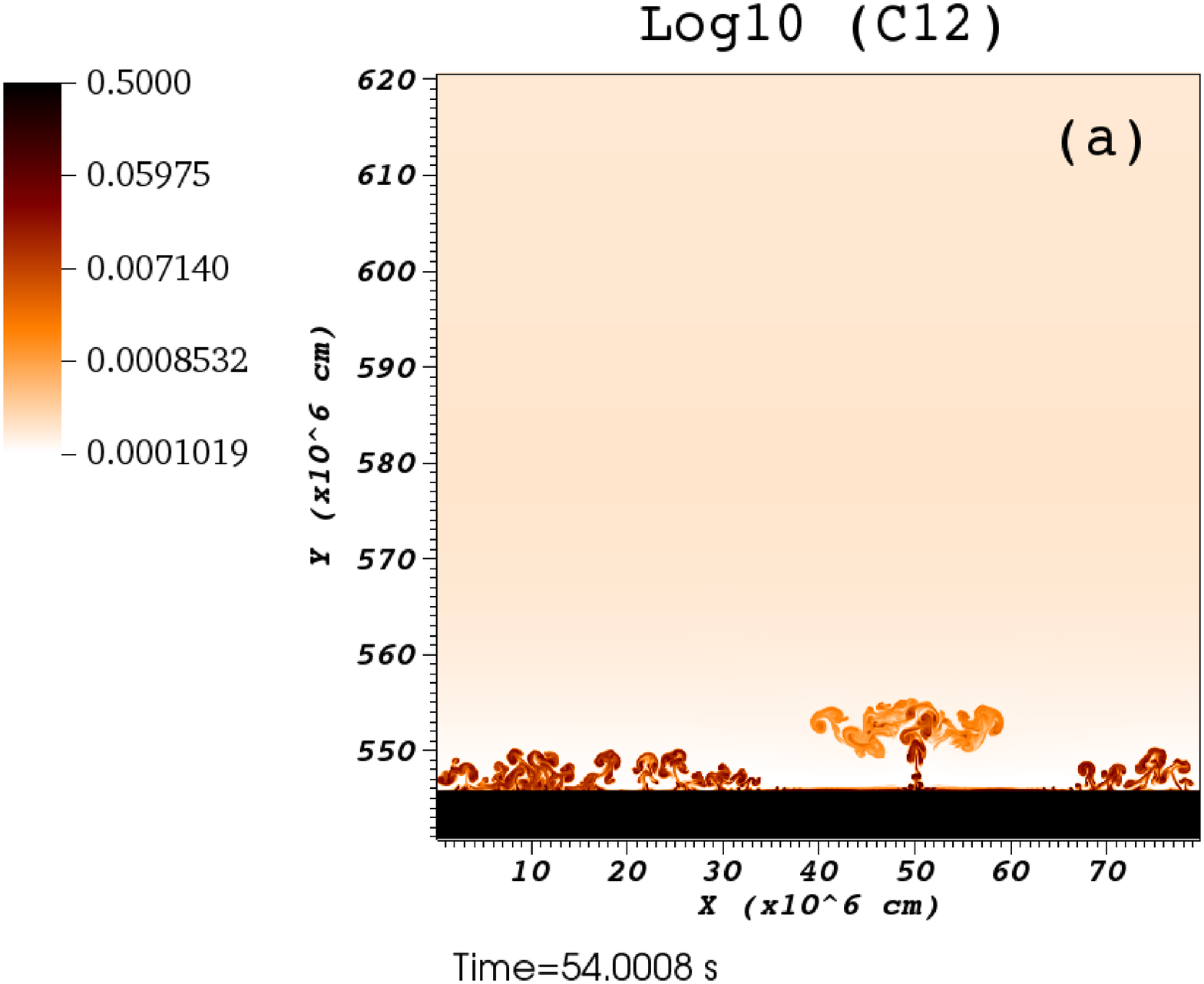}
   \includegraphics[width=0.475\textwidth,height=0.3\textheight]{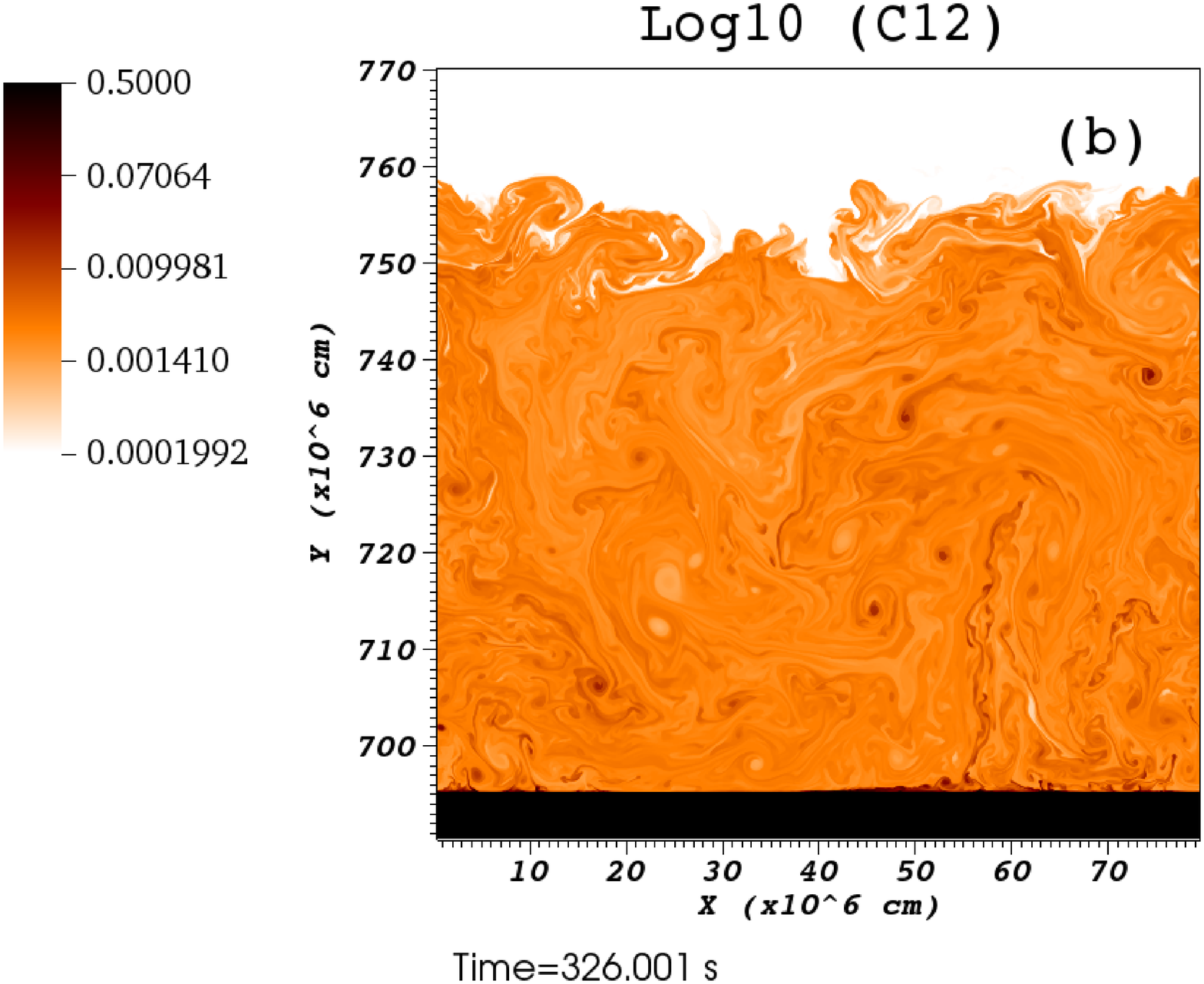}
   \includegraphics[width=0.475\textwidth,height=0.3\textheight]{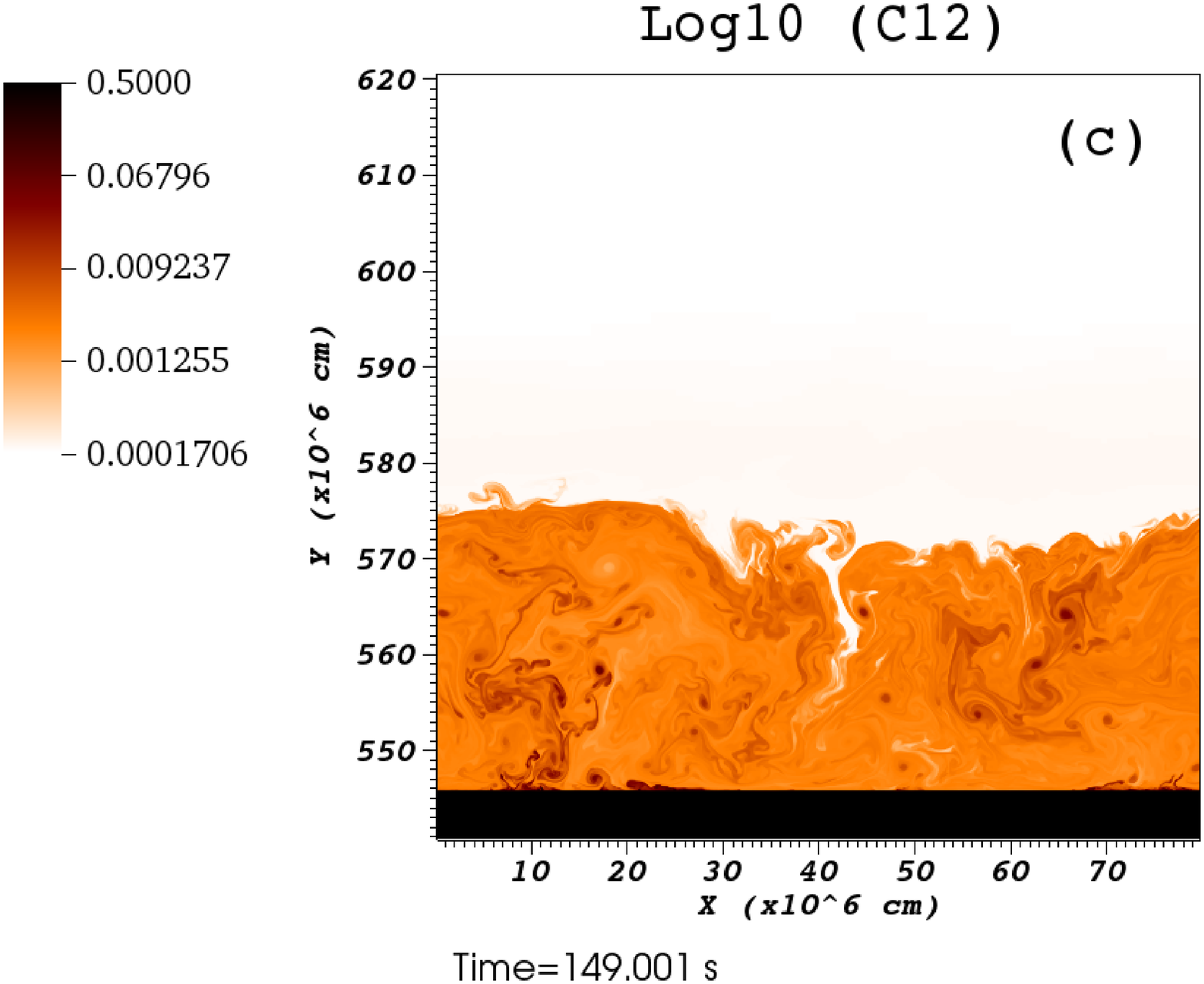}
   \includegraphics[width=0.475\textwidth,height=0.3\textheight]{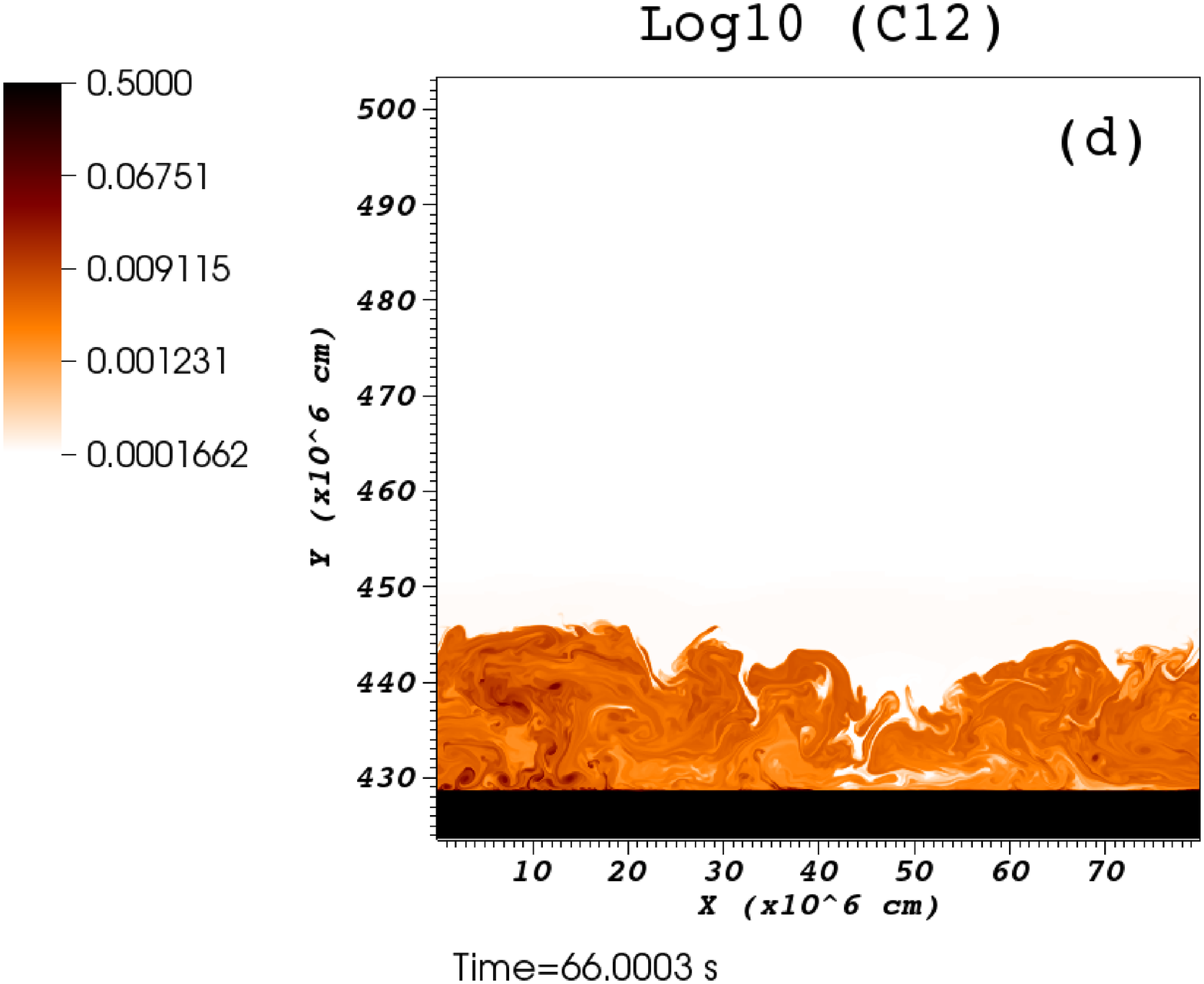}
   \includegraphics[width=0.475\textwidth,height=0.3\textheight]{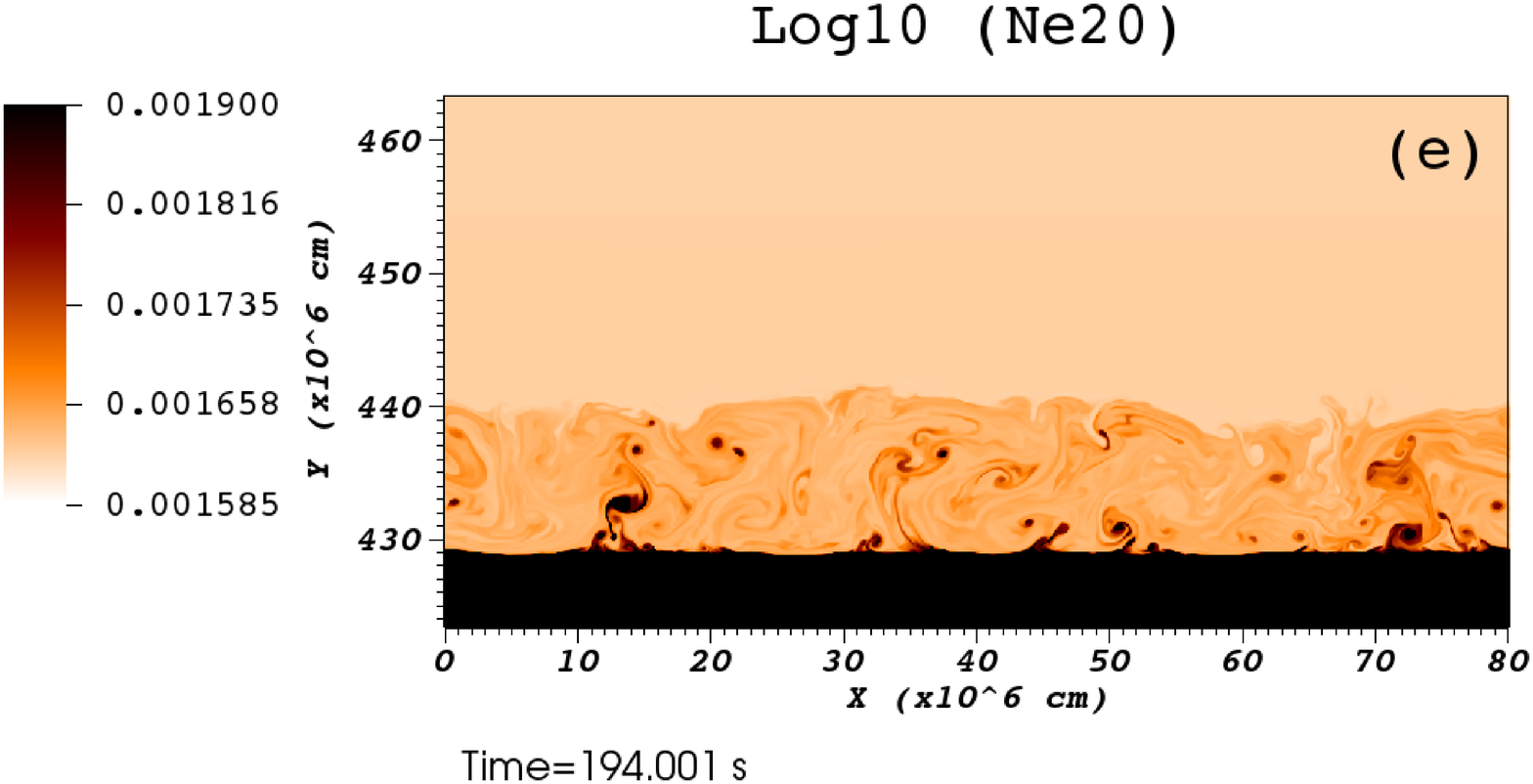}
   \includegraphics[width=0.475\textwidth,height=0.3\textheight]{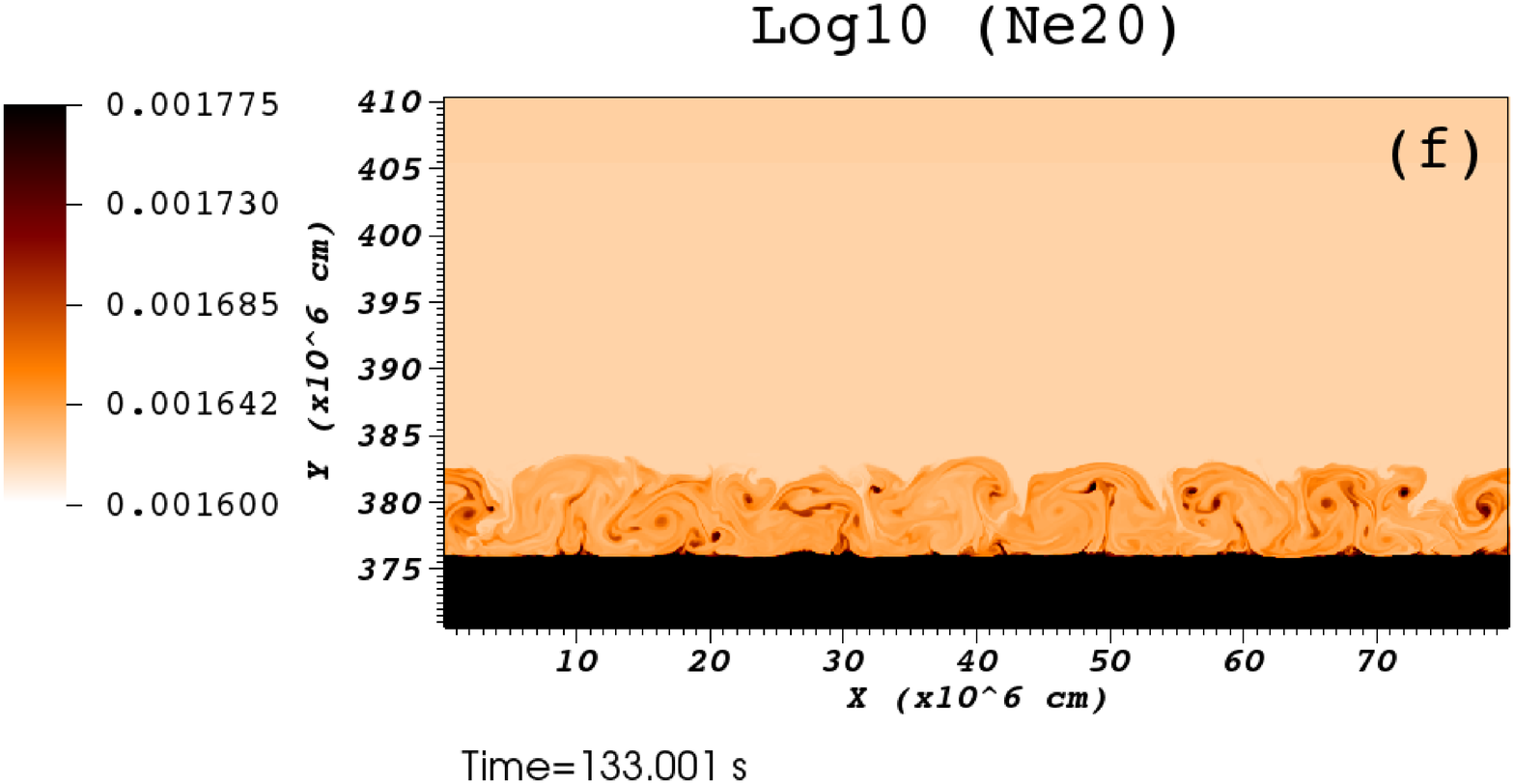}

       \caption{Panel (a) shows the growth of the primary fluid instabilities at t=54 s, in terms of $^{12}$C mass fraction and in logarithmic scale, for model
                1.0 $M_{\odot}$ CO white dwarf. The other panels show snapshots of the convective front when it is at a distance of 4 pressure scale heights
                above the CEI for models (b) 0.8 $M_{\odot}$ CO, (c) 1.0 $M_{\odot}$ CO, (d) 1.15 $M_{\odot}$ CO, (e) 1.15 $M_{\odot}$ ONe,
                 and (f) 1.25 $M_{\odot}$ ONe. The snapshots are shown in terms of $^{12}$C mass fraction for CO-rich substrates, and in terms of $^{20}$Ne mass fraction for ONe-rich substrates, both in logarithmic scale. The times are t=326 s, t=149 s, t=66 s, t=194 s, and t=133 s, respectively. The reflection condition is imposed in the lower boundary, which is located 50 km below the CEI for all the simulations. See movies.
              }
         \label{FigFront}
   \end{figure*}
%

   \begin{figure*}
   \includegraphics[width=0.35\textwidth,height=0.35\textheight,angle=270]{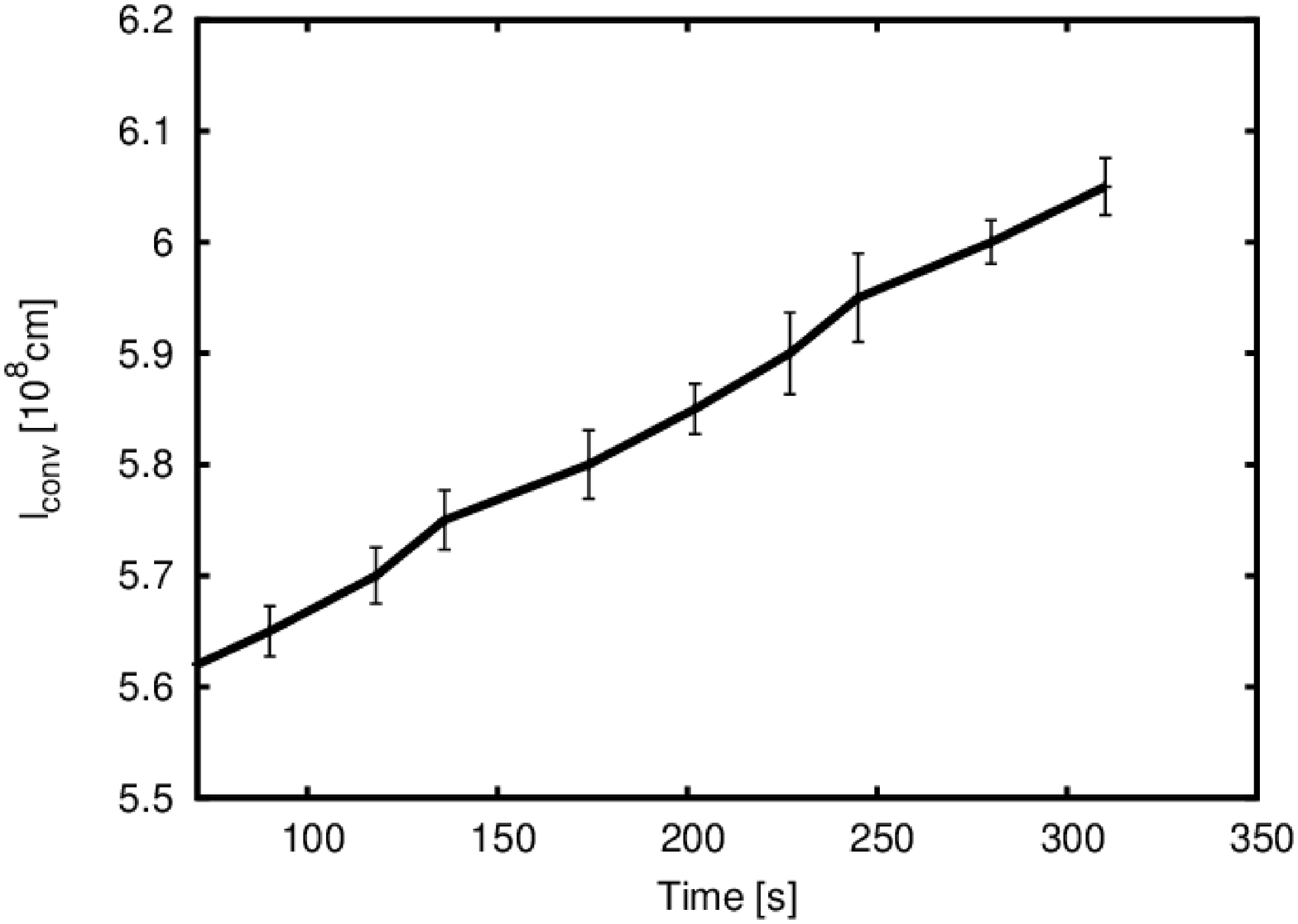}
   \includegraphics[width=0.35\textwidth,height=0.35\textheight,angle=270]{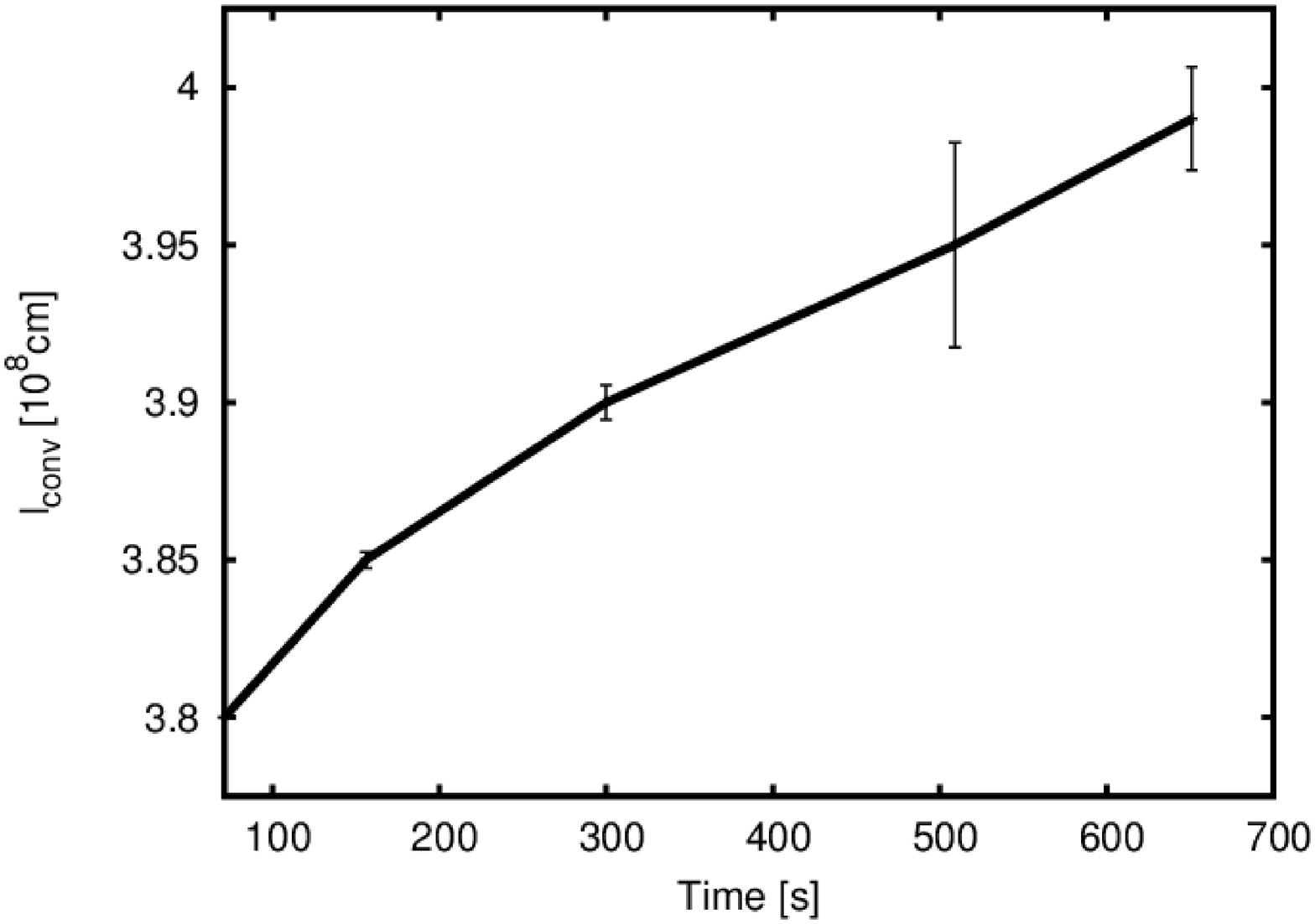}
       \caption{Propagation of the convective front throughout the accreted envelope as a function of time, for models 1.0 $M_{\odot}$ CO (panel a) and 1.25 $M_{\odot}$ ONe (panel b).
              }
         \label{FigVibStab}
   \end{figure*}
%

   \begin{figure*}
   \includegraphics[width=0.35\textwidth,height=0.35\textheight,angle=270]{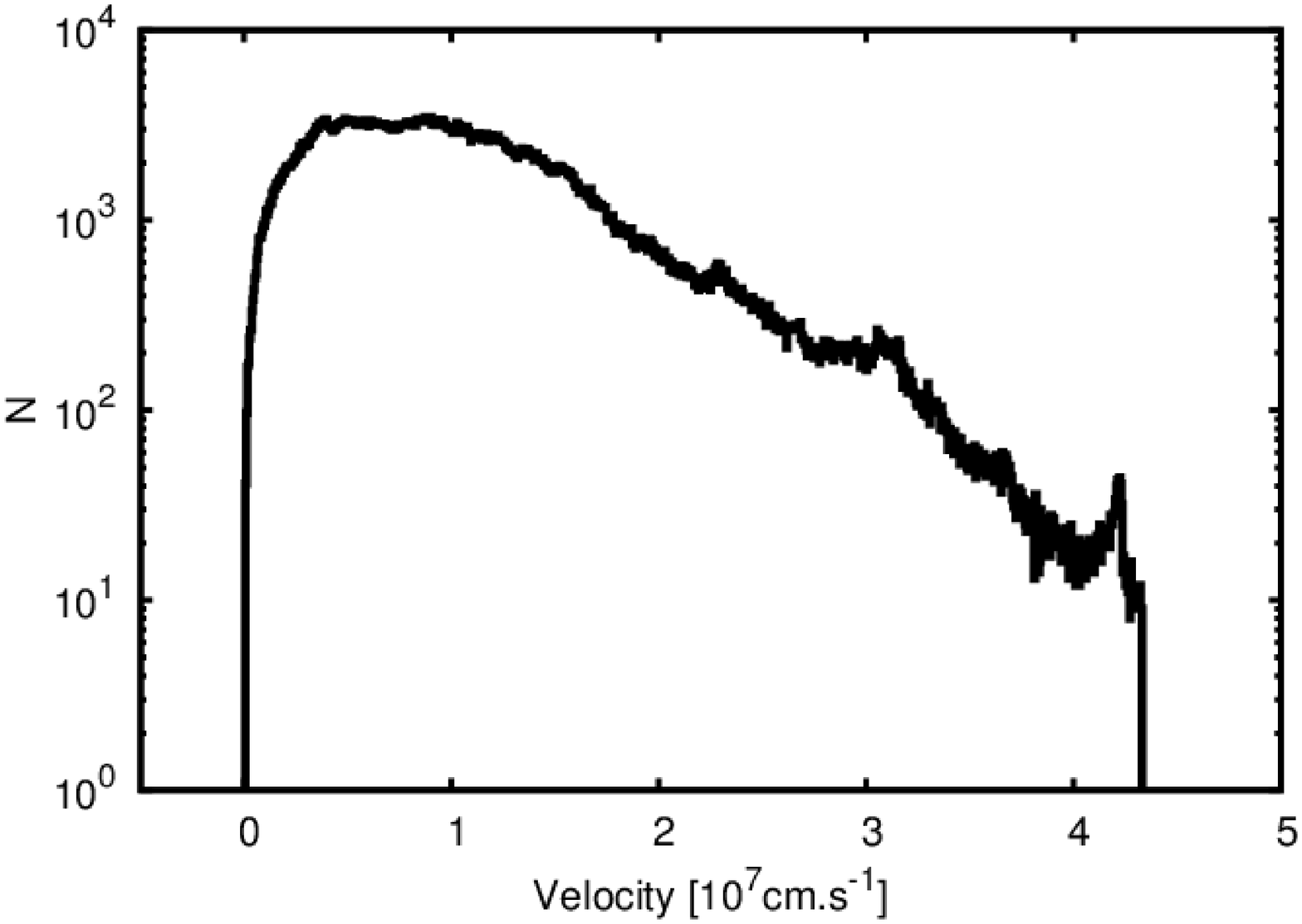}
   \includegraphics[width=0.35\textwidth,height=0.35\textheight,angle=270]{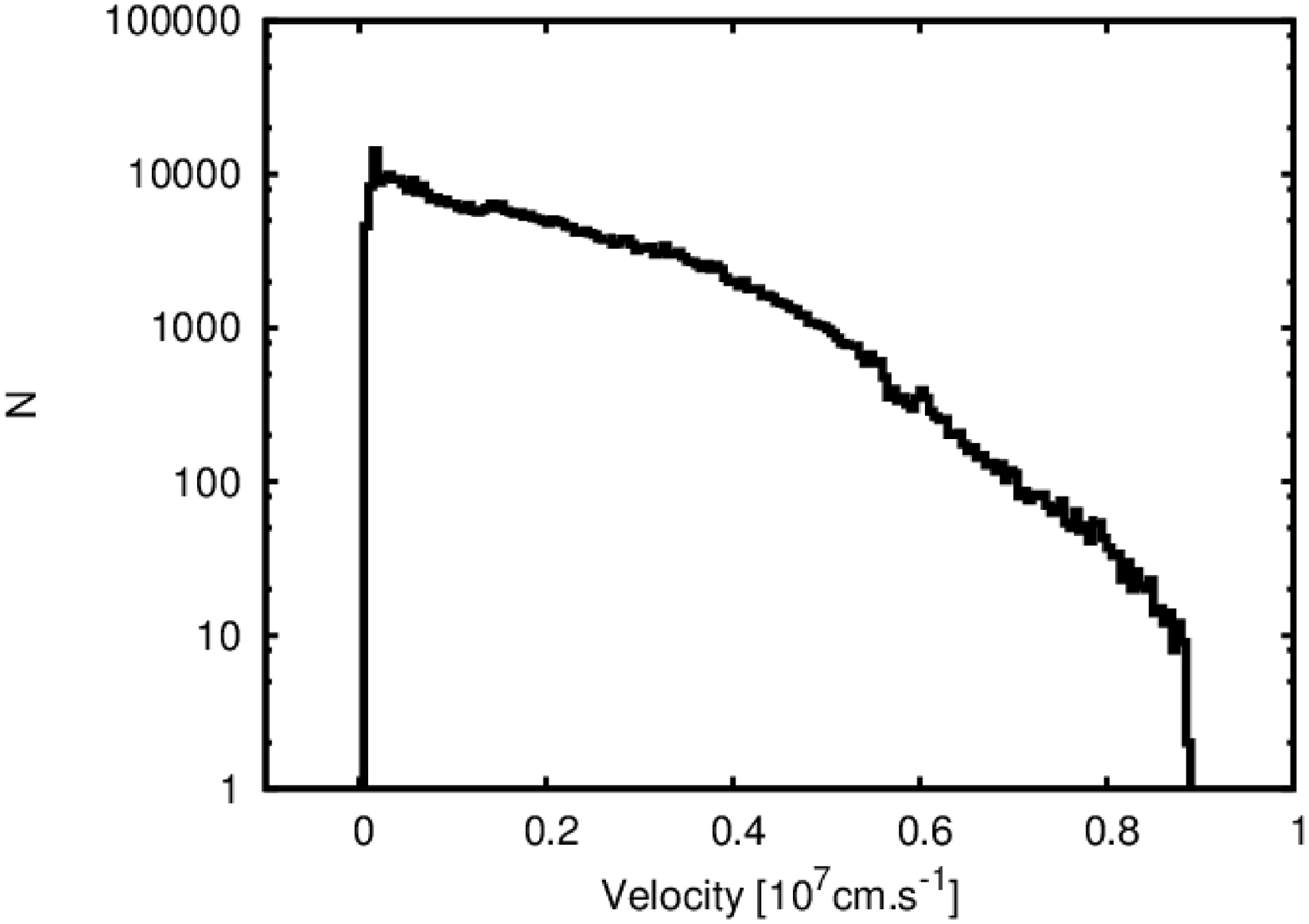}
       \caption{Velocity histograms for models 1.0 $M_{\odot}$ CO at t=310 s (panel a) and 1.25 $M_{\odot}$ ONe at t=657 s (panel b). The discontinuities in the velocity are due to different stages of the turbulent mixing during the burning advance.
              }
         \label{FigVibStab}
   \end{figure*}

   \begin{figure*}
   \includegraphics[width=0.35\textwidth,height=0.35\textheight,angle=270]{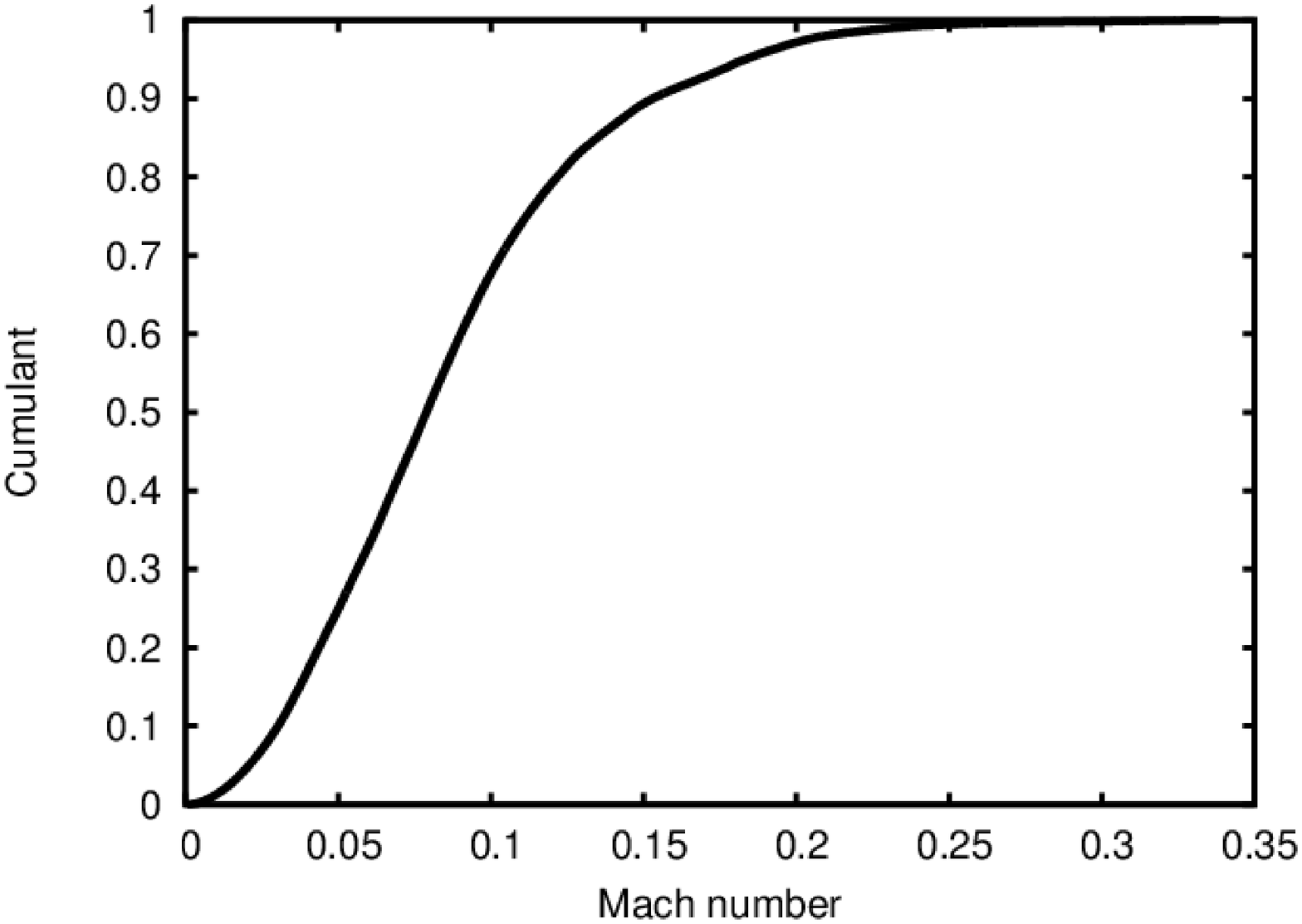}
   \includegraphics[width=0.35\textwidth,height=0.35\textheight,angle=270]{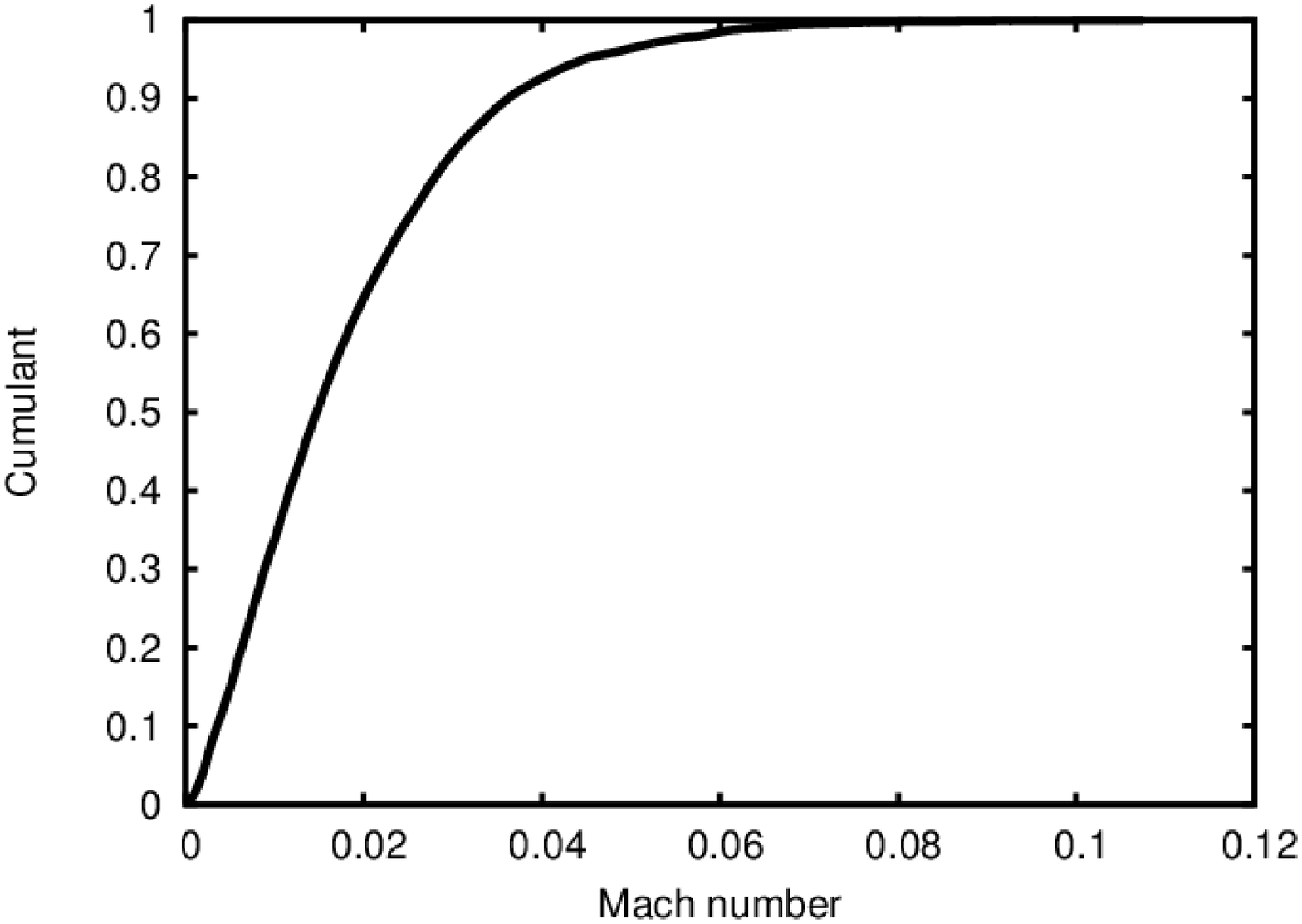}
       \caption{Cumulant distribution function of the Mach number for models 1.0 $M_{\odot}$ CO at t=310 s (panel a) and 1.25 $M_{\odot}$ ONe at t=657 s (panel b). The cumulant distribution shows the probability that the Mach number is equal or below a certain value.
              }
         \label{FigVibStab}
   \end{figure*}

   \begin{figure*}
   \includegraphics[width=0.35\textwidth,height=0.35\textheight,angle=270]{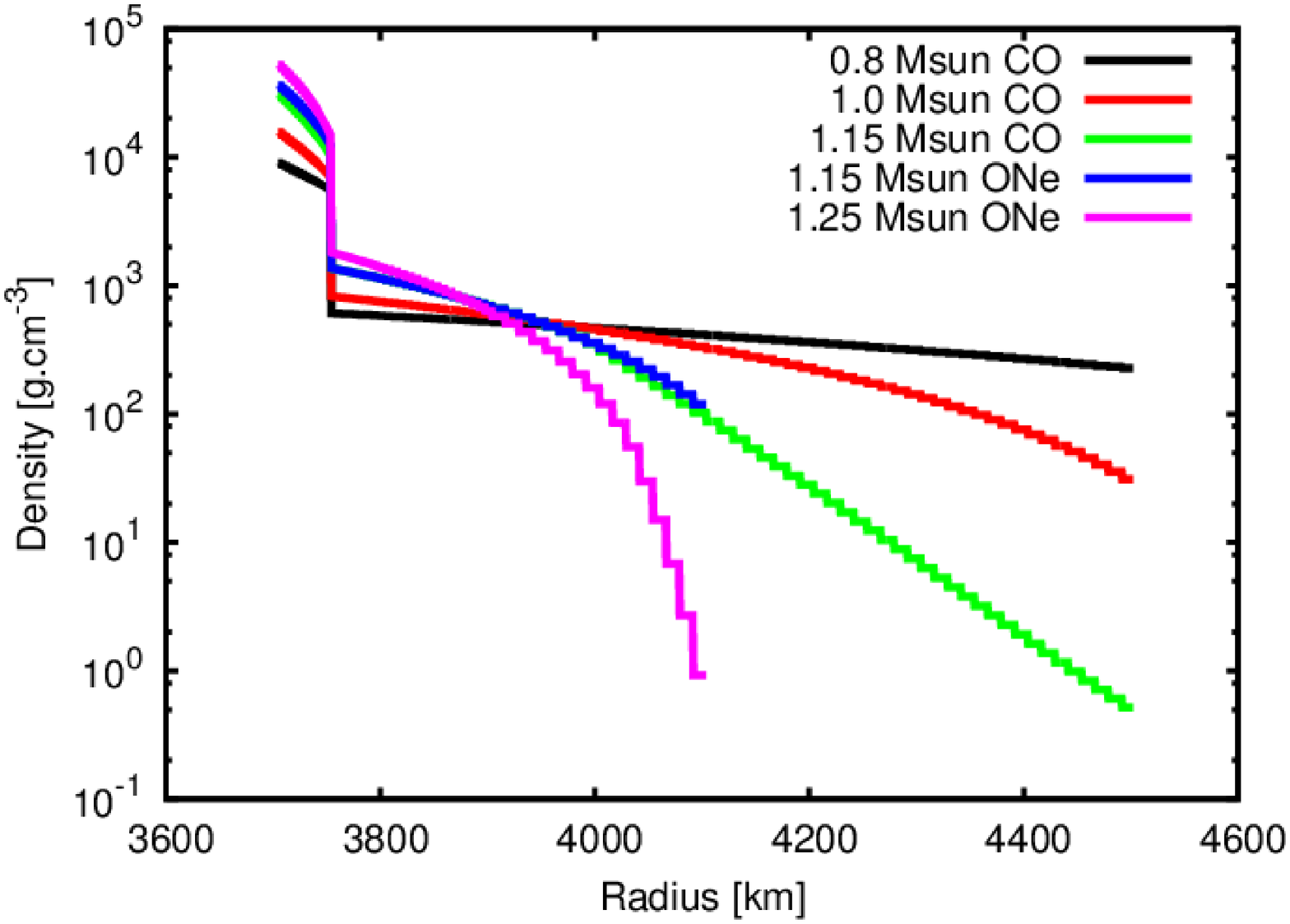}
   \includegraphics[width=0.35\textwidth,height=0.35\textheight,angle=270]{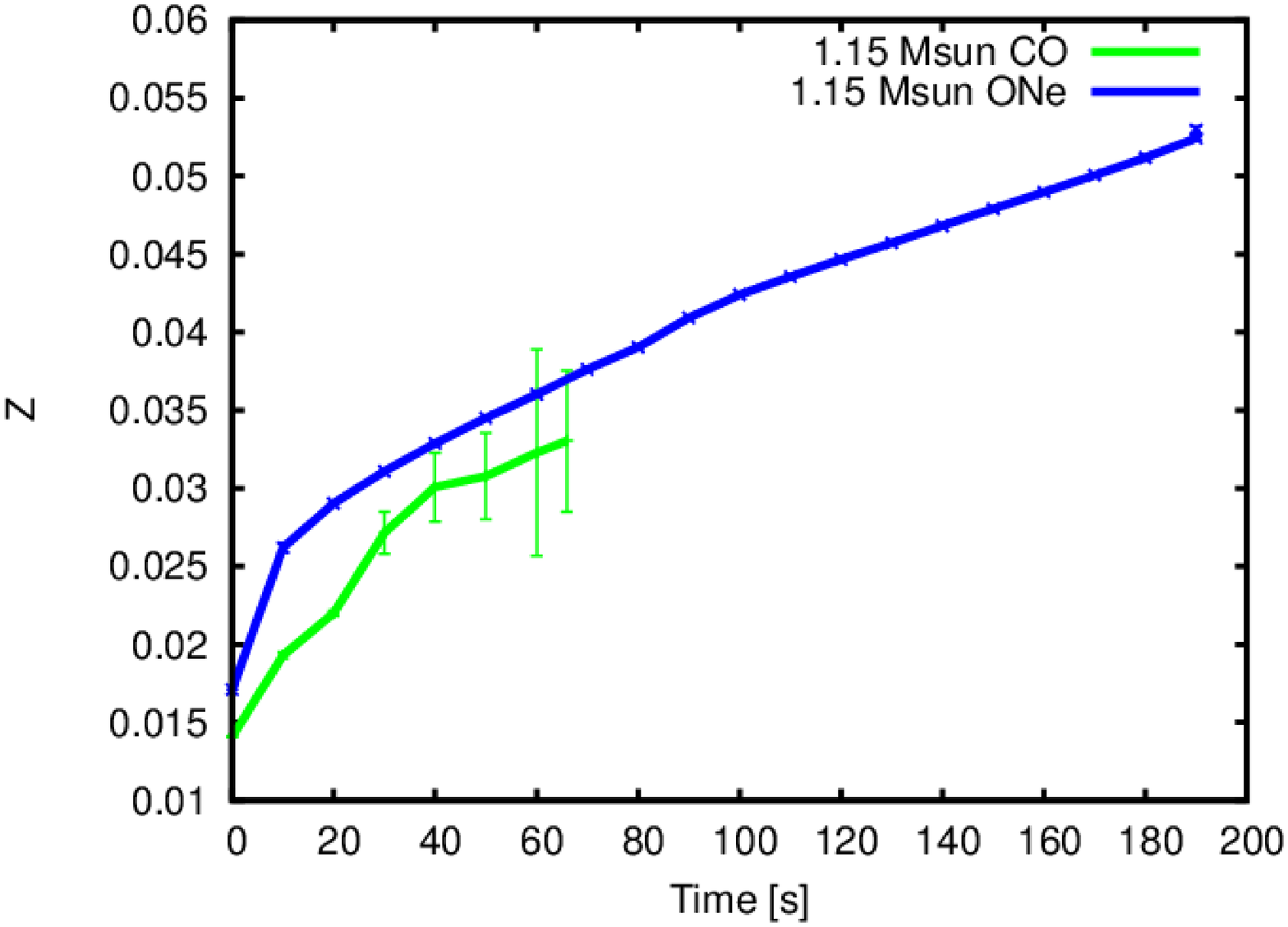}
       \caption{Panel a: initial density profiles for the five computed models. Panel b: Comparison of the time evolution of the mean metallicity throughout the envelope for models 1.15 $M_{\odot}$ CO and 1.15 $M_{\odot}$ ONe until the convective front reaches 4 pressure scale heights above the CEI.
              }
         \label{FigVibStab}
   \end{figure*}
%

%
\begin{table*}
\caption{Main charactheristics of the computed models}
\label{table:1}
\centering
\begin{tabular}{c c c c c c c c c c c }     
\hline\hline
Nova & Size (km$\times$km)  & $M_{wd}$ ($M_{\odot}$) & g (cm.s$^{-2}$) & Hp (km) & d (km)  & $t_{inst}$ (s) & $t_{Hp}$ (s) & Z & $T_{base}$ (K)\\
\hline
   CO &  800$\times$800 & 0.8 & -2.20$\times$10$^{8}$ & 156.8 & 627.2  & 60 & 326 & 0.0264 & 1.12$\times$10$^{8}$  \\
   CO &  800$\times$800 & 1.0 & -4.46$\times$10$^{8}$ & 71.2  & 284.6  & 40    & 149 & 0.0295 & 1.10$\times$10$^{8}$ \\
   CO &  800$\times$800 & 1.15 & -8.32$\times$10$^{8}$ & 36.7 & 146.9 & 20     & 66 & 0.0330 & 1.07$\times$10$^{8}$ \\
   ONe & 800$\times$400  & 1.15 & -8.32$\times$10$^{8}$ & 27.6 & 110.5 & 5    & 194  & 0.0530 & 1.01$\times$10$^{8}$ \\
   ONe & 800$\times$400  & 1.25 & -1.18$\times$10$^{9}$ & 16.6 & 66.4  & 4    & 133 & 0.0652 & 1.01$\times$10$^{8}$ \\
\hline
\end{tabular}
\end{table*}
%

   \begin{figure*}
   \includegraphics[width=0.35\textwidth,height=0.35\textheight,angle=270]{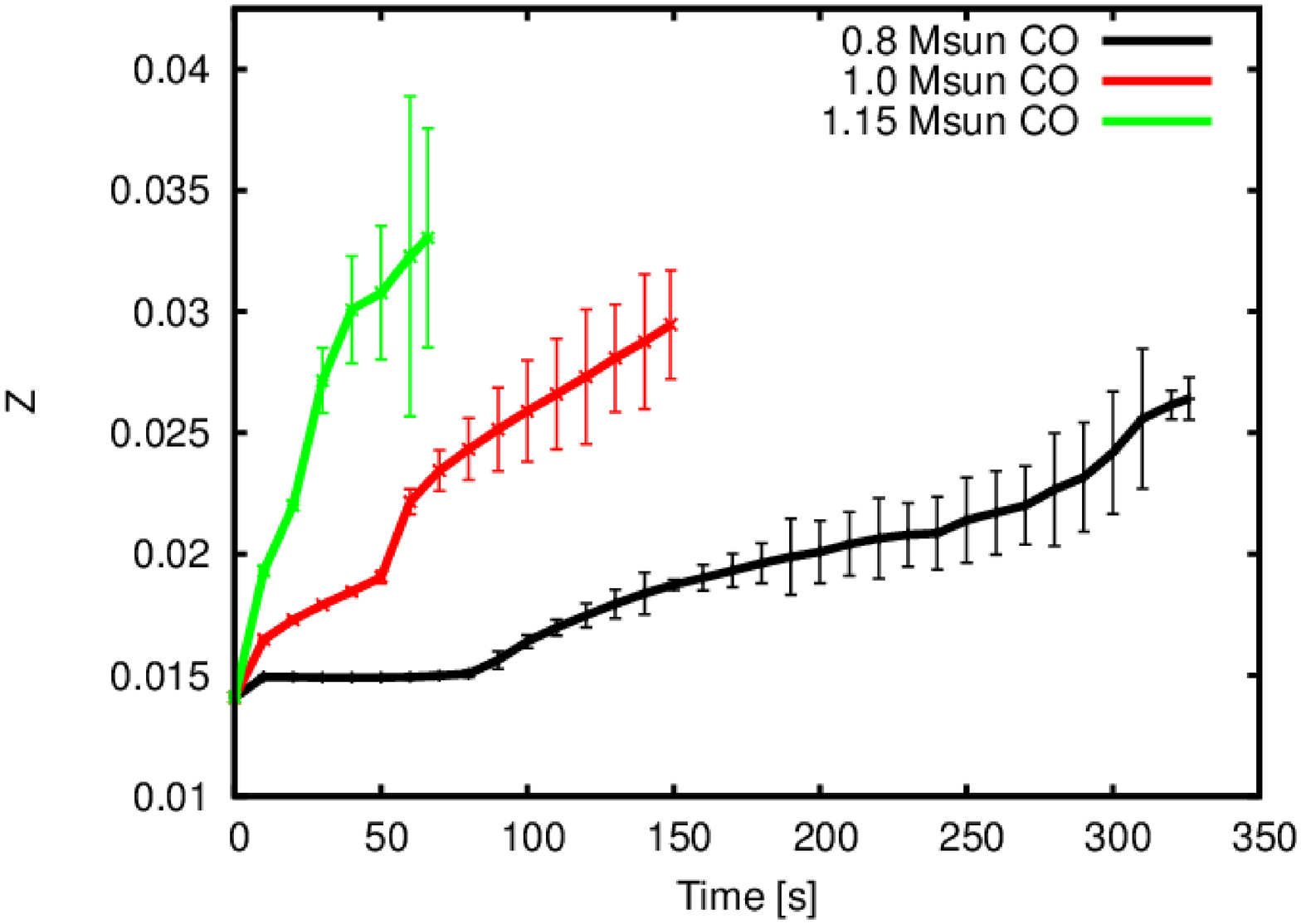}
   \includegraphics[width=0.35\textwidth,height=0.35\textheight,angle=270]{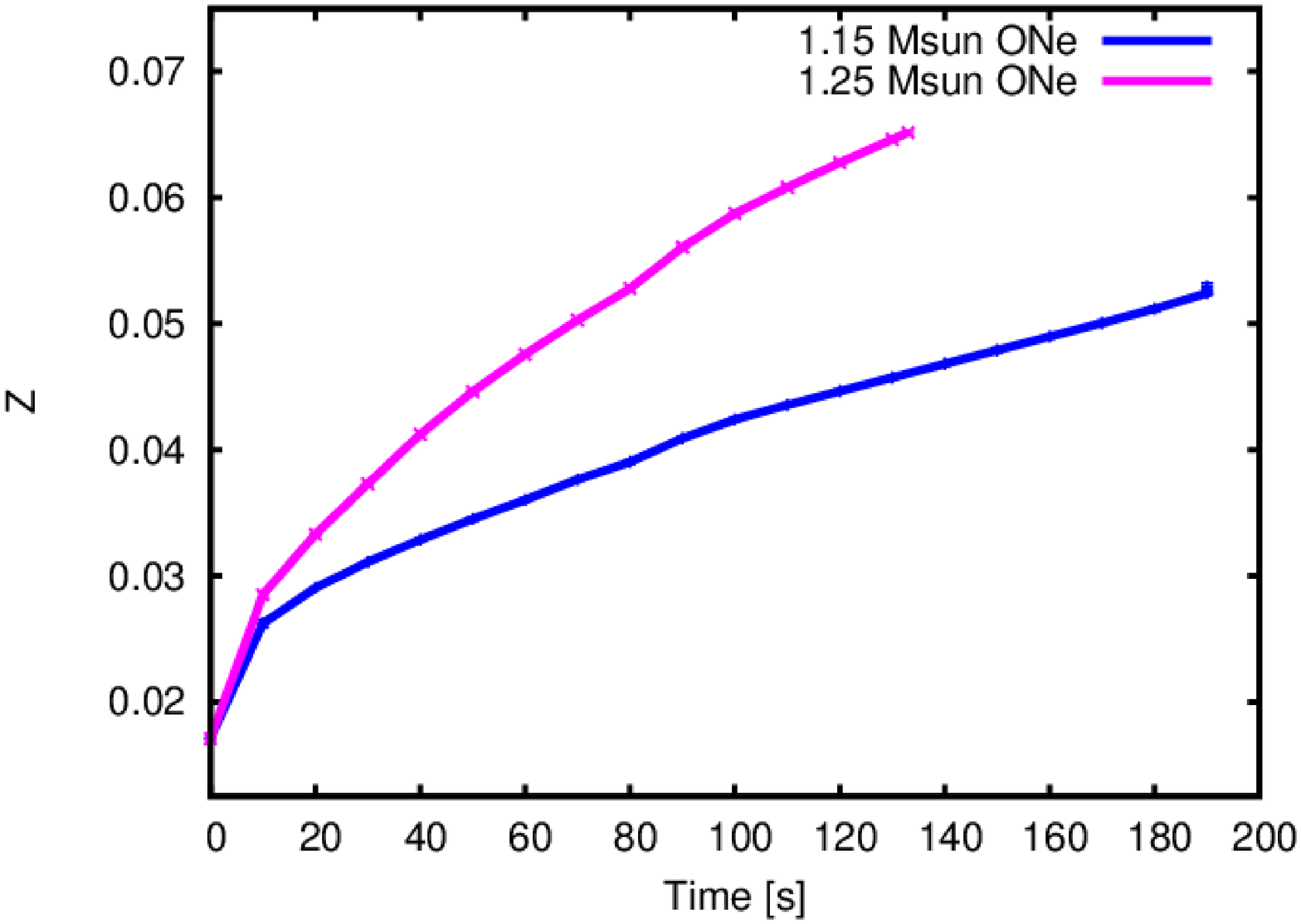}

       \caption{Evolution of the mean metallicity with time in CO-rich (panel a) and ONe-rich (panel b) substrates at 4 pressure scale heights above the CEI, respectively. More massive white dwarfs result in larger metallicity enhancements.
              }
         \label{FigVibStab}
   \end{figure*}

   \begin{figure*}
   \includegraphics[width=0.35\textwidth,height=0.35\textheight,angle=270]{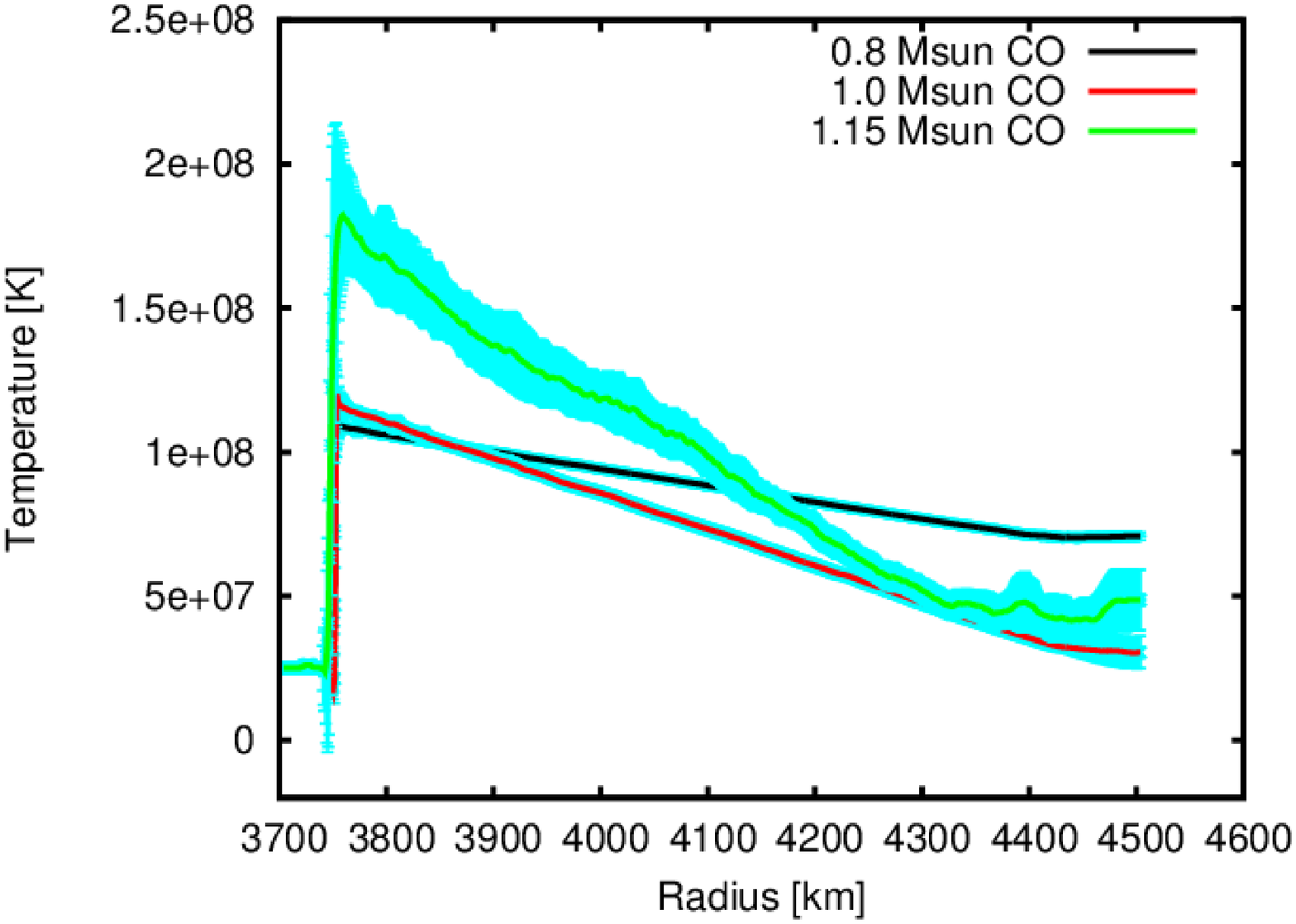}
   \includegraphics[width=0.35\textwidth,height=0.35\textheight,angle=270]{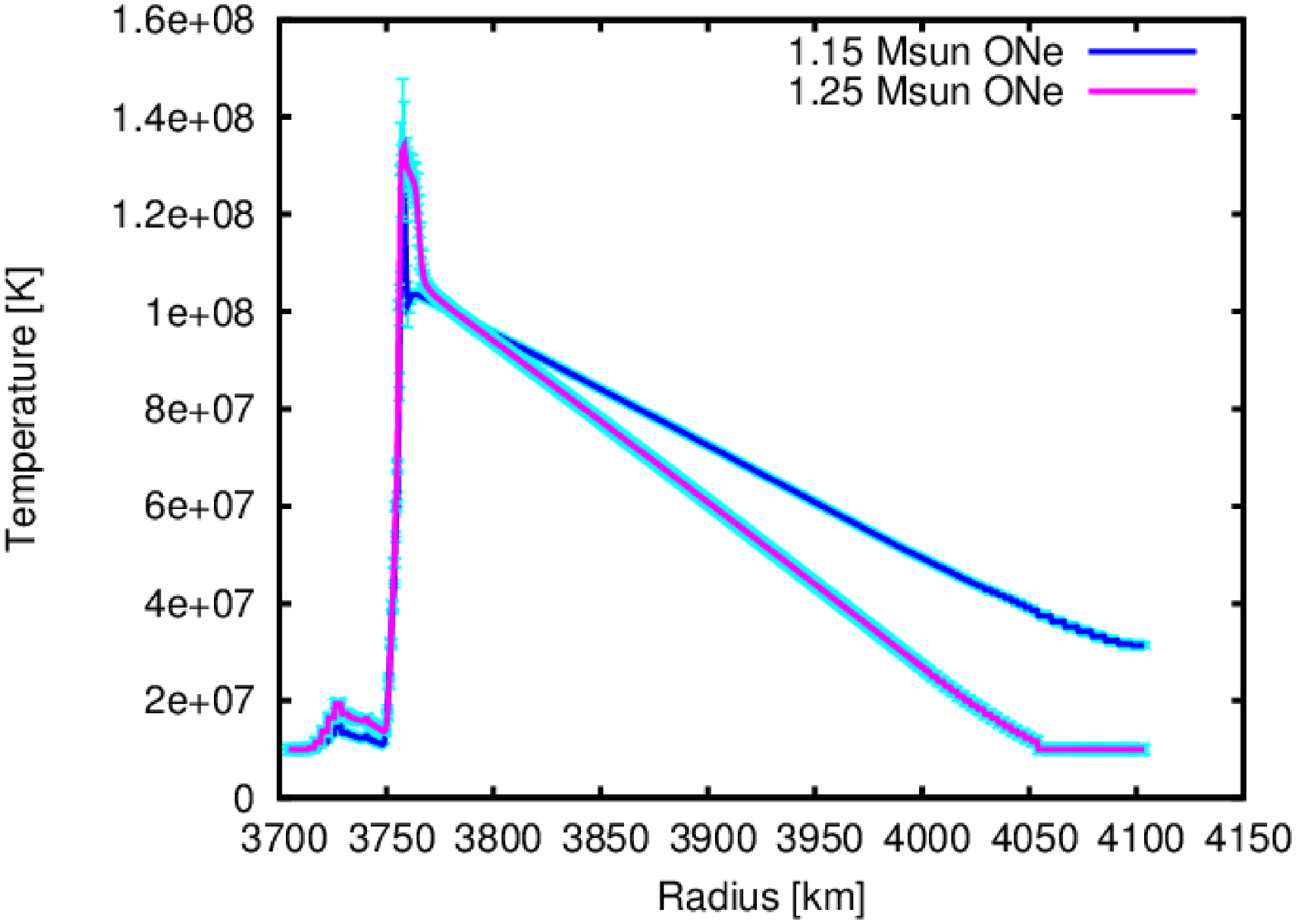}

       \caption{Averaged radial temperature profile in CO-rich (panel a) and ONe-rich (panel b) substrates at the time the convective front almost reaches the upper boundary, 650 km and 250 km above the CEI, respectively.
                Base temperatures are higher in more massive white dwarfs due to higher pressures at the base of the envelope.
              }
         \label{FigVibStab}
   \end{figure*}

   \begin{figure*}
   \includegraphics[width=0.475\textwidth,height=0.3\textheight]{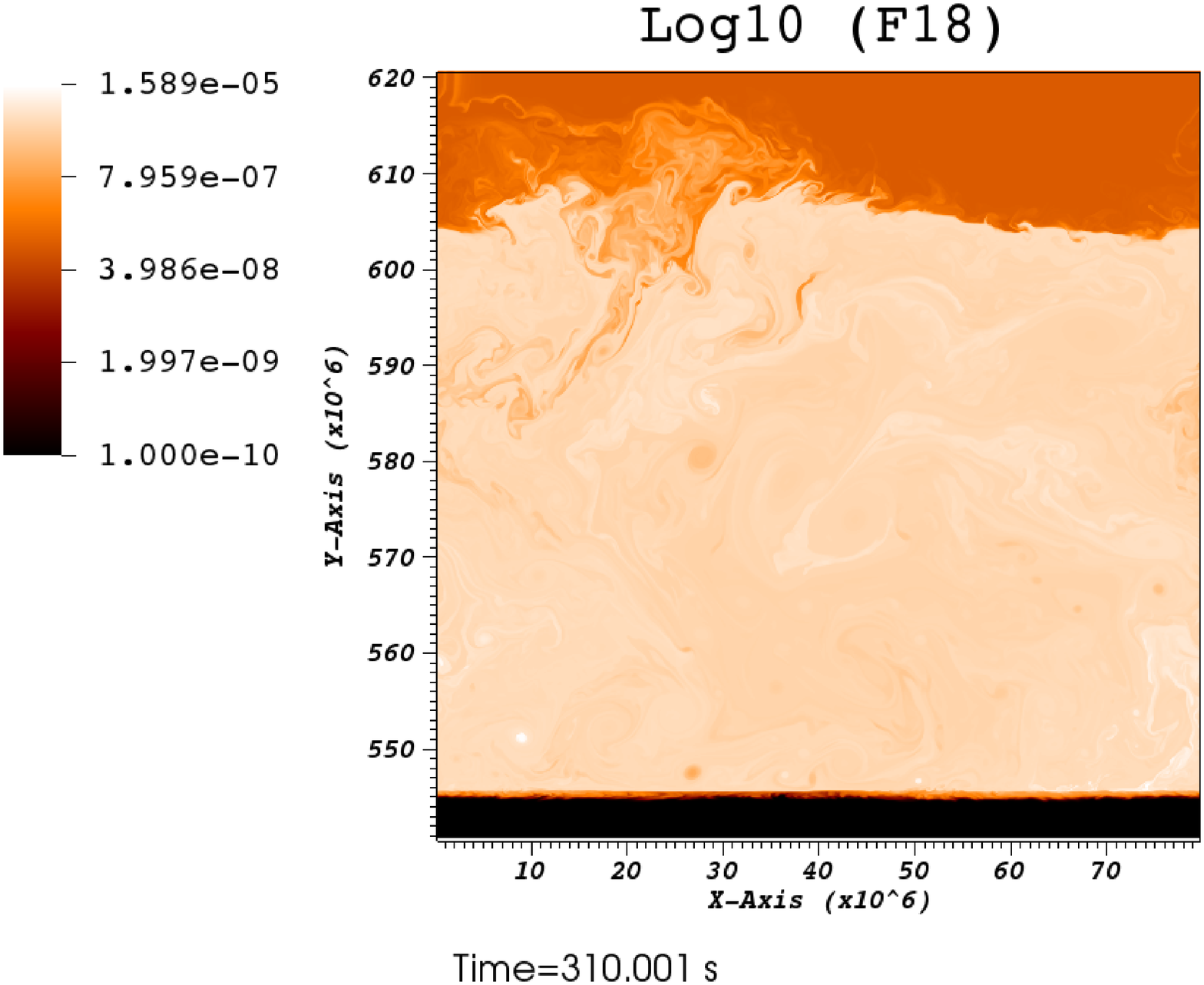}
   \includegraphics[width=0.475\textwidth,height=0.3\textheight]{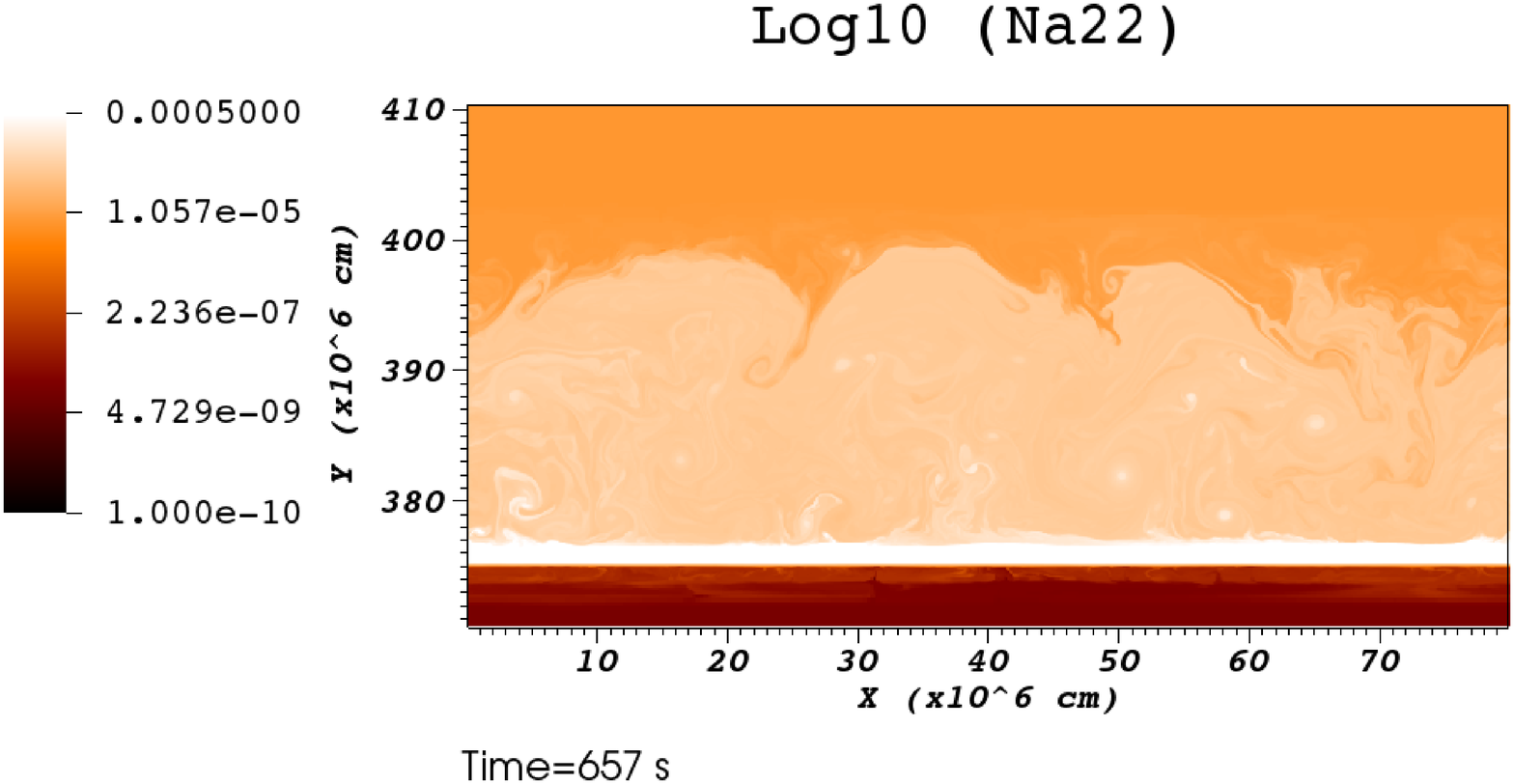}

       \caption{Development of the initial fluid instabilities at the CEI at t=310 s, in terms of $^{18}$F mass fraction and in logarithmic scale, for model 1.0 $M_{\odot}$ CO (panel a), and at t=657 s, in terms of $^{22}$Na mass fraction and in logarithmic scale, for model 1.25 $M_{\odot}$ ONe (panel b). Identical plots are obtained for the other chemical species.
              }
         \label{FigVibStab}
   \end{figure*}

   \begin{figure*}
   \includegraphics[width=0.35\textwidth,height=0.35\textheight,angle=270]{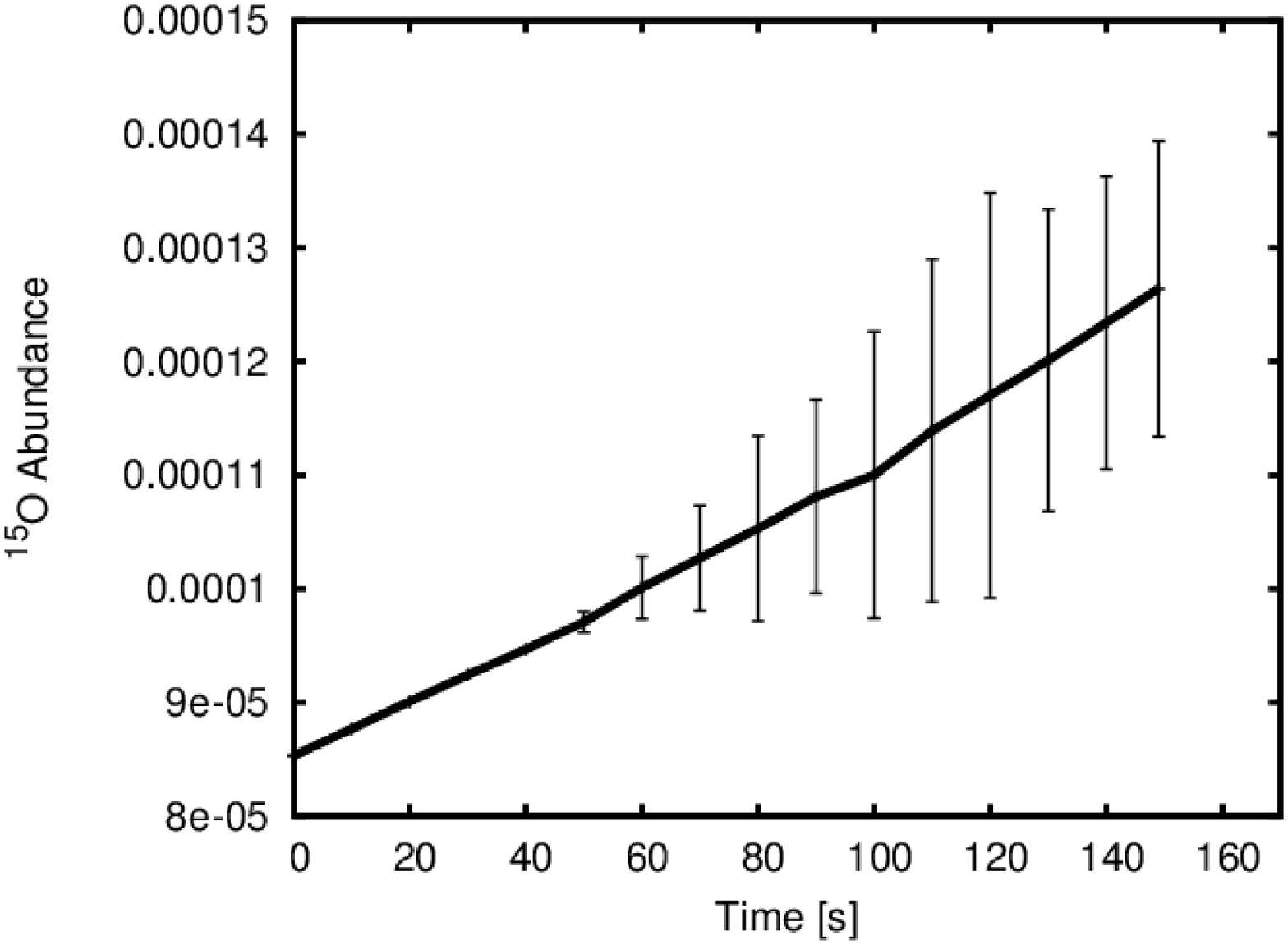}
   \includegraphics[width=0.35\textwidth,height=0.35\textheight,angle=270]{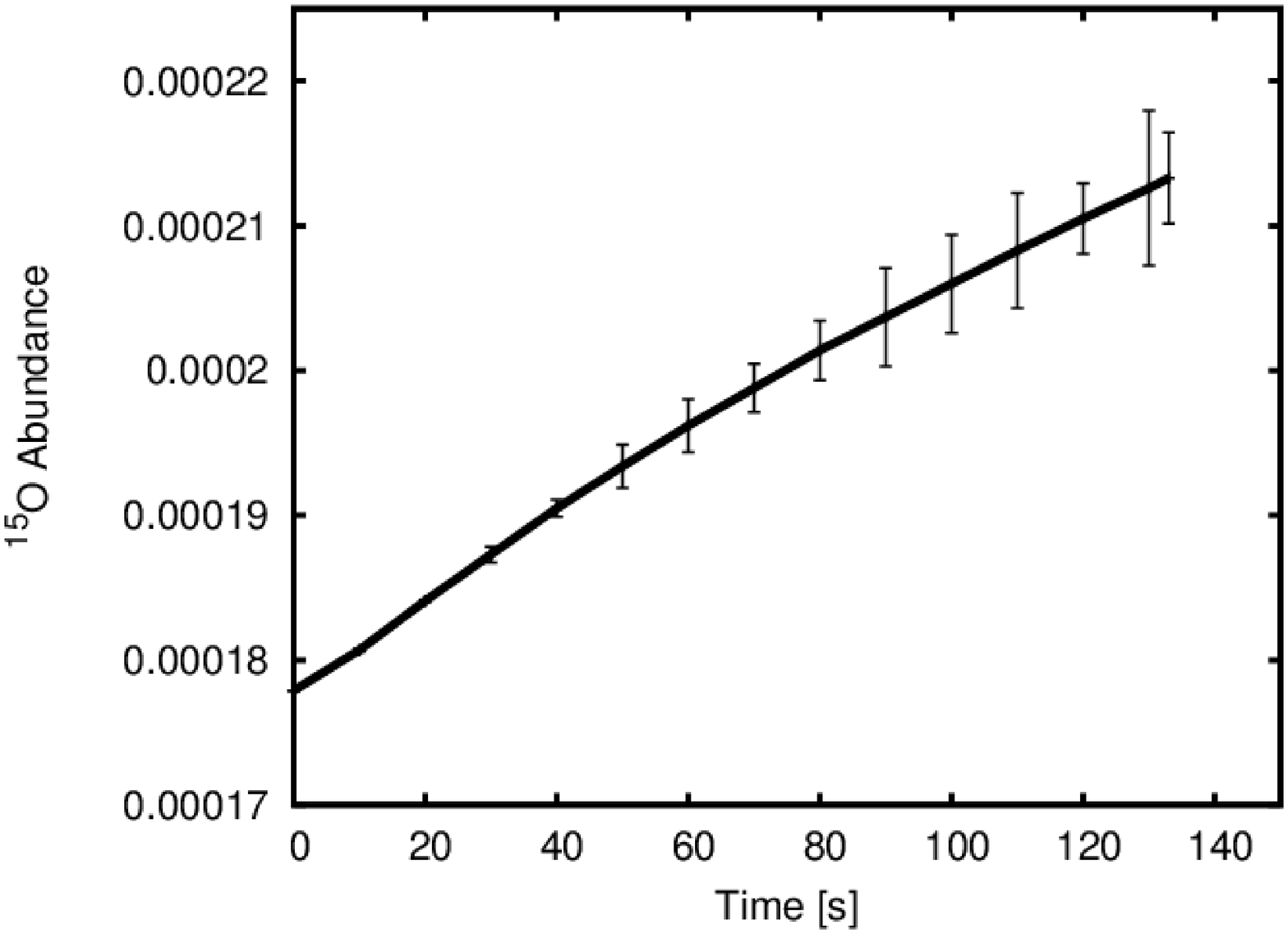}
       \caption{Evolution of the mean $^{15}$O abundance with time throughout the envelope until the convective front reaches 4 pressure scale heights, for models 1.0 $M_{\odot}$ CO (panel a) and 1.25 $M_{\odot}$ ONe (panel b).
              }
         \label{FigVibStab}
   \end{figure*}

\section{Results}
We initiated the simulations by introducing a top-hat temperature perturbation (5$\%$) operating only at the first timestep in a 1 km wide region located near the CEI. The location of the perturbation is the same for the five computed models. Previous work \citep{Cas11a} has shown that the outcome is insensitive to the nature of the perturbation. The perturbation disrupts the initial equilibrium configuration and creates fluctuations at the interface that spawn strong buoyant fingers. In Fig. 1, panel (a) shows the development of these primary fluctuations at t=54 s for the 1.0 $M_{\odot}$ CO model. 
Panels (b), (c), (d), (e), and (f) in Fig. 1 show the progress of the convective front at the time it is located at 4 pressure scale heights above the CEI, for the different models reported in this work. The scale pressure height, at the times indicated in Fig. 1, is calculated as H$_{p}$=kT/mg, where k is the Boltzmann constant, T is the temperature, m is the molecular mass, and g is the value of the gravity. The two 1.15 $M_{sun}$ models have the same gravity, but once mixing operates, the values of T and m differ, resulting in different values of H$_{p}$. Four pressure scale heights corresponds to 627.2 km, 284.6 km, 146.9 km, 110.5 km, and 66.4 km above the CEI for models 0.8 $M_{\odot}$ CO, 1.0 $M_{\odot}$ CO, 1.15 $M_{\odot}$ CO, 1.15 $M_{\odot}$ ONe, and 1.25 $M_{\odot}$ ONe, respectively. 
Shear flows develop at the interface and trigger strong Kelvin-Helmholtz instabilities that efficiently dredge-up core material and enrich the envelope in metals.The convective front progresses subsonically upwards almost uniformly, such that the event proceeds almost in spherical symmetry. That explains the success of one-dimensional calculations in describing the main characteristics of classical nova explosions \citep{Star98,Kov97,Yaron05,Jose98}. 

The advance of the convective front is limited by the abundance stratification and the temperature gradient within the accreted envelope. Consequently, matter burns layer by layer depending on the composition and the temperature that characterizes each stratum, and the front moves upwards almost uniformly. Fig. 2 demonstrates the advance of the convective front as a function of time until the convective front has nearly reached the upper boundary at t=310 s and t=657 s, for models 1.0 $M_{\odot}$ CO and 1.25 $M_{\odot}$ ONe, respectively.  
As burning proceeds, convection becomes more disordered\footnote{Based on the prescription by \citet{Spit62} as implemented in the FLASH code, typical values of the Reynolds and Prandtl numbers 
are $\sim$ 10$^{19}$and $\sim$ 10$^{-7}$, respectively.}, and the 2D convective cells continue growing in size, occupying almost the entire domain length. As opposed to the 3D case \citep{Cas11b,Cas16}, conservation of vorticity in 2D forces eddies to grow in an inverse vorticity cascade with a distribution that deviates from the Kolmogorov spectrum \citep{Pope00,Lei08,Shore07}. In a 3D framework, the fluid shows a more chaotic pattern, where filaments originating at the CEI tend to break into smaller structures while being engulfed at the upper parts of the envelope. The formation of long-lived eddies of different sizes is a signature of the intermittent behavior of turbulence. Movies showing the development of the Kelvin-Helmoltz instabilities, and the progress of the convective front for models 1.0 $M_{\odot}$ CO and 1.25 $M_{\odot}$ ONe are available online\footnote{or at http://www.fen.upc.edu/users/jjose/Downloads.html}. Notice that the movie of the 1.25 $M_{\odot}$ ONe shows a Landau-like bifurcation at $\sim$ 100 s \citep{Land87}. This bifurcation appears when non-linear modes become strong enough, that the fluid becomes unstable and develops an oscillatory transition into a more tangled regime. In Fig. 3, we show the total velocity histograms at the end of the simulation for model 1.0 $M_{\odot}$ CO at t=310 s (mean velocity=1.06$\times$10$^{7}$cm s$^{-1}$ and dispersion=6.74$\times$10$^{6}$cm s $^{-1}$), and model 1.25 $M_{\odot}$ ONe at t=657 s (mean velocity=1.87$\times$10$^{6}$cm s$^{-1}$ and dispersion=1.48$\times$10$^{6}$cm s$^{-1}$). Model 1.0 $M_{\odot}$ CO presents a more developed turbulent regime that translates into a distribution that is approximately Gaussian, with higher mean velocity and a lower dispersion. The 1.25 $M_{\odot}$ ONe model instead, presents a distribution with a larger dispersion and lower velocities, indicating that turbulence is less developed. The same trend is found in the cumulant distribution functions of the Mach number for these models (see Fig. 4). Mean values of the Mach numbers and their corresponding variances are: $\bar{Ma}$=0.0862 and $\sigma^{2}$=0.0002 for model 1.0 $M_{\odot}$ CO, and $\bar{Ma}$=0.0180 and $\sigma^{2}$=0.0024 for model 1.25 $M_{\odot}$ ONe. The inhomogenous distribution with an extended tail, is distinctly non-Gaussian, indicating that these quantities present a signature of the intermittent behavior of the turbulence \citep{Cas11b}.
 
When the convective front reached 4 pressure scale heights above the interface, we obtain reference parameters for the five computed models. In Table 1, we list the main characteristics at this time: white dwarf substrate (Nova), size of the computational domain (Size), mass of the white dwarf ($M_{wd}$), the value of the gravity at the CEI (g), rise time of the initial buoyant fingering ($t_{inst}$)\footnote{$t_{inst}$ corresponds to the time at which the first instability arises at the CEI for each computed model.}, value of the pressure scale height (Hp), distance above the CEI at which the convective front is at (d), final computed time at which the convective front is at 4 pressure scale heights above the CEI ($t_{Hp}$), mean metallicity throughout the accreted envelope (Z), and base temperature ($T_{base}$). The initial temperature perturbation, that diffuses rapidly, creates hot spots that burn faster and transfer energy to the surrounding material, which becomes buoyant. The initial buoyant fingering appears sooner in the simulations for ONe-rich substrates and for more massive white dwarfs due to higher density values in the accreted envelope and steeper density gradients which, in turn, translate into a higher pressure at the base (see $t_{inst}$ in Table 1, and panel (a) in Fig. 5) \citep{Jose16}. The buoyant fingering time for the ONe models and the formation of small convective eddies in the innermost layers of the envelope is delayed relative to the CO models. For instance, in the 1.25 $M_{\odot}$ ONe model, it takes $\sim$ 35 s for the initial buoyant fingering to develop convection, while it only takes $\sim$ 20 s in the 1.0 $M_{\odot}$ CO model (see movies). Rising plumes contain hot gas that is replaced by material from the cold upper layers, and effective circulation initiates at the innermost layers of the accreted envelope. However, when this occurs depends on the ignition timescale. For solar material\footnote{See, however, \citet{Shen09} for ignition conditions in C-poor envelopes.}, ignition is driven by the reaction $^{12}$C+p, which is faster than other channels, such as $^{16}$O+p or $^{20}$Ne+p \citep{Jose16b}. Hence, ignition and the establishment of superadiabatic gradients required for convection occur earlier in CO-rich substrates\footnote{Notice that superadiabatic gradients are not a requirement for buoyancy.}.
In ONe-rich substrates, the amount of $^{12}$C that can be excavated from the core by Kelvin-Helmholtz instabilities is small compared to CO-rich substrates (X(C$^{12}$)=0.00916 and X(C$^{12}$)=0.5, respectively), which results in a longer dredge-up episode, and therefore, a higher degree of mixing \citep{Cas16}. The timescale of the explosion (as well as the completness of convection) is, consequently, longer in ONe-rich substrates, requiring more than 600 s to reach the upper boundary, as opposed to $\sim$ 300 s in CO-rich substrates (see movies)\footnote{It is worth mentioning that the extent of the accreted envelope is twice as large as those of the CO-rich substrate models.}. Notice also that the convective front propagates faster in CO-rich substrates (see Fig. 2). In Fig. 5, panel (b) shows a comparison of the evolution of the metallicity between models 1.15 $M_{\odot}$ CO and 1.15 $M_{\odot}$ ONe at the time the convective front reached 4 pressure scale heights.
The mean metallicities are indeed higher at the time we stop the simulations, $\sim$ 30$\%$, in ONe-rich substrates (Z=$0.0330\pm0.0045$ in CO-rich substrates vs Z=$0.0530\pm0.0003$ in ONe-rich substrates). We observe a much more homogeneous metallicity enhancement for ONe models (notice the smaller dispersion that these models present in panel (b) of Fig. 5). Mixing is less uniform in CO-rich substrates because a larger amount of $^{12}C$ can be dredged up from the core. Therefore, burning proceeds faster, resulting in a more developed turbulent regime, which in turn, translates into a more inhomogeneous filamentary behavior. The trend observed in the final metallicities is also found in spherical (1D) models \citep{Jose98}.

It is worth mentioning that the burning front for these models is the boundary of the convecting 
region. Its advance is the consequence of the time development of the 
nuclear source and the nonlinear process of thermally buoyant motions 
inducing dynamical instabilities. We cannot capture the fully developed 
stage because for the relatively short duration of the simulations, the 
relatively limited size of the computational domain, and the restriction 
of all motions to two dimensions. But notwithstanding these limitations 
the models suffice to make some firm statements. There is a cascade, to 
the extent that a broad spectrum of motion and density clearly develops,
such that the models will lead to a broad hierarchy of mixing with large coherent structures 
persisting that differ in chemical composition. 

In Fig. 6, we show the evolution of the metallicity throughout the envelope with time for CO-rich (panel (a)) and ONe-rich substrates (panel (b)) until the convective front reaches 4 pressure scale heights. Since the base density and peak temperatures are higher for more massive white dwarfs, the reaction rates increase, which translates into substantial nuclear processing that involves heavier elements. More massive white dwarfs result in a higher metallicity enhancement. For instance, in CO-rich substrates, the final metallicities at the time we stop the simulations are Z=0.0268 for 0.8 $M_{\odot}$, Z=0.0772 for 1.0 $M_{\odot}$, and Z=0.2651 for 1.15 $M_{\odot}$. In ONe-rich substrates, we obtain Z=0.0866 for 1.15 $M_{\odot}$, and Z=0.1174 for 1.25 $M_{\odot}$. These final mean metallicities are higher than those reported in Table 1 because they are measured at the time the convective front almost reached the upper boundary ($\sim$ 650 km above the CEI for CO models, and $\sim$ 250 km for ONe models)\footnote{Final times are t=333 s for model 0.8 $M_{\odot}$ CO, t=310 s for model 1.0 $M_{\odot}$ CO, t=218 s for model 1.15 $M_{\odot}$ CO, t=667 s for model 1.15 $M_{\odot}$ ONe, and t=657 s for model 1.25 $M_{\odot}$ ONe.}. 

Peak temperatures are determined by the pressure achieved at the base of the envelope before ignition. Accordingly, more massive white dwarfs reach higher pressures and higher peak temperatures (see radial temperature profiles, measured at the time the convective front reached the upper boundary, in panels (a) and (b) of Fig. 7). Base temperatures in Fig. 7 are higher than those reported in Table 1 because these are measured at the end of the simulation. At the time we stop the simulations, base temperatures still keep rising, but we expect them to level off when detachment results. The maximum temperatures (the peak value is not yet reached) are lower in ONe-rich substrates at the time we stop the simulations, although we would also expect them to be higher for ONe models than for CO models at the time of detachment \citep{Jose98}. In ONe-rich substrates, the nuclear activity extends beyond the CNO mass region, shifting towards the MgAl and NeNa regions where heavier isotopes are produced (see the development of fluid instabilities in terms of $^{18}$F and $^{22}$Na mass fractions depending on the adopted underlying white dwarf in Fig. 8). In Fig. 9 we show the evolution of the $^{15}$O abundance for models 1.0 $M_{\odot}$ CO (panel a) and 1.25 $M_{\odot}$ ONe (panel b). The dispersion found in ONe models is again smaller than for CO models, indicating a more uniform enhancement.

\section{Discussion}
In this paper, we have explored the viability of Kelvin-Helmholtz instabilities operating at the CEI as a mechanism for self-enrichment of the accreted envelope with material from the underlying white dwarf, using different substrates and white dwarf masses. While this was partially investigated by \citet{Glasner12}, in the framework of CO, pure He, pure $^{16}$O, and pure $^{24}$Mg substrates, our work adopts a more realistic composition for an ONe-rich substrate, and extends the analysis for a range of white dwarf masses. All the initial models have been self-consistently computed (in 1D), and hence, the adopted envelope masses are also more realistic. The 2-D simulations reported in this paper show that ONe-rich substrates produce a longer dredge-up episode and higher resulting metallicities within the accreted envelope. Finally, we also find more mixing and more energetic explosions for more massive white dwarfs. 

A detailed analysis of the development of turbulence in 2-D and 3-D frameworks in classical nova explosions and characterization of its intermittent behavior is currently being performed.
It is worth noting that the convective pattern found in our simulations is an unvoidable consequence of the assumed two dimensional configuration, forcing the fluid motion to behave differently than in 3-D convection \citep{Shore07,Meakin07}. In 2-D, the convective cells are forced to merge into large convective eddies with a size comparable to the envelope height. Nevertheless, the levels of metallicity enhancements and the essential explosion properties are recovered in a 2-D framework, since fully 3-D simulations differ qualitatively, but not quantitatively, from 2-D simulations \citep{Cas11b}. 

The different envelope size in our 2-D simulations, although imposed by the properties of the underlying white dwarf, makes it hard to quantify the final envelope metallicities. The metallicity values reported here represent lower limits, since the simulations were stopped when the convective front reaches the top of the computational domain and much before the expected detachment of the layers occurs. Extended models are necessary to simulate the entire nova outburst from mixing to the final expansion of the envelope, and to rigorously quantify the dredge-up and investigate the progress of the explosion and subsequent ejection. Since our goal is to investigate the mixing mechanism during the thermonuclear runaway, we start the simulations from 1D models that assume no pre-mixing. However, forcing the initial 1D model to be in hydrostatic equilibrium after mapping into 2-D imposes a restriction on the portion of the envelope that can be computed, since the density dramatically decreases and produces numerical underflows. A procedure to expand the initial models under hydrostatic equilibrium was explored by \citet{Zingale02}. These authors stopped placing the initial model into hydrostatic equilibrium when a certain density cutoff was reached and added a portion of matter with very low constant density on top of it. This method was found to have a negligible impact on their results. We plan to investigate a similar procedure to extend the initial models in future calculations

A final aspect needs further discussion. In our 1D solar accretion models, the envelope is fully convective when the base temperature is $\sim$ 9 $\times$ $10^{7}$ K. A realistic comparison between CO and ONe models should be made when the rates for proton capture on $^{12}$C and on $^{20}$Ne are similar. At the time we map our models (base temperatures of $10^{8}$ K), the proton capture rate on $^{12}$C is much faster than that on $^{20}$Ne. Therefore, CO models should be mapped at earlier times to be consistent with ONe models, regarding the length of the mixing process. Previous studies in 2D \citep{Glasner97,Glasner07} investigated how mapping the initial 1D models at lower temperatures ($\sim$ 7 $\times$ $10^{7}$ K) influences the overall mixing. However, the authors found a minor impact on the results.

\begin{acknowledgements}
      Part of the software used in this work was developed by the DOE-supposed ASC/Alliances Center for Astrophysical Thermonuclear Flashes at the University of Chicago. This work has been partially supported by the Spanish MINECO grant AYA2017-86274-P, by the E.U. FEDER funds, by the AGAUR/Generalitat de Catalunya grant SGR-661/2017, and by the U.S. Department of Energy, Office of Science, Office of Nuclear Physics. We also acknowledge the Barcelona Supercomputing Center for a generous allocation of time at the MareNostrum supercomputer. This article benefited from discussions within the "ChETEC COST Action (CA16117). 
\end{acknowledgements}

%
%

\end{document}